\documentclass[aps,twocolumn,nofootinbib,preprintnumbers,groupedaddress,letterpaper]{revtex4-2}

\usepackage{hyperref}

\usepackage[dvipsnames]{xcolor}
\usepackage[normalem]{ulem}
\usepackage{amsmath}
\usepackage{enumerate}
\usepackage{mathtools}

\usepackage{amsfonts}
\usepackage{amssymb}

\usepackage[colorinlistoftodos,prependcaption,textsize=tiny]{todonotes}

\usepackage{graphicx}

\usepackage{lipsum}

\usepackage{enumitem}
\usepackage{pifont}
\def\teller{P\"{o}schl-Teller }

\def\psyk{\mathsf{p}}

\newcommand\comment[1]{}

\newcommand\schr{Schr\" odinger }

\newcommand\poincare{Poincar\' e }

\newcommand\ov{\over }

\def\le{\left}
\def\ri{\right}

\def\({\left(}
\def\){\right)}
\def\<{\langle}
\def\>{\rangle}

\newcommand\half{{\ensuremath{\frac{1}{2}}}}
\newcommand\p{\ensuremath{\partial}}

\newcommand\field[1]{{\ensuremath{\mathbb{{#1}}}}}

\newcommand{\RR}{\field{R}}

\newcommand{\be}{\begin{equation}}
\newcommand{\ee}{\end{equation}}
\newcommand{\bea}{\begin{eqnarray}}
\newcommand{\eea}{\end{eqnarray}}
\newcommand{\bwt}{\begin{widetext}}
\newcommand{\ewt}{\end{widetext}}

\newcommand{\bi}{\begin{itemize}}
\newcommand{\ei}{\end{itemize}}
\newcommand{\ben}{\begin{enumerate}}
\newcommand{\een}{\end{enumerate}}
\newcommand{\bca}{\begin{cases}}
\newcommand{\eca}{\end{cases}}
\newcommand{\bln}{\begin{align}}
\newcommand{\eln}{\end{align}}
\newcommand{\bst}{\begin{split}}
\newcommand{\est}{\end{split}}

\def\XXint#1#2#3{{\setbox0=\hbox{$#1{#2#3}{\int}$}
\vcenter{\hbox{$#2#3$}}\kern-.5\wd0}}

\begin{document}

\preprint{QMUL-PH-25-25}

\title{A folded string dual for the Sachdev--Ye--Kitaev model}

\author{David Vegh}
\email{d.vegh@qmul.ac.uk}

\affiliation{\it  Centre for Theoretical Physics, Department of Physics and Astronomy,
Queen Mary University of London, 327 Mile End Road, London E1 4NS, UK}

\begin{abstract}

We propose a folded string moving in rigid AdS$_2$ with {\it imaginary radius squared} as a dual of the Sachdev--Ye--Kitaev (SYK) model at its conformal fixed point. In standard AdS$_2$, the string is represented by two massless particles connected by straight string segments. The particles move at the speed of light, abruptly reversing direction at turning points. We describe the system using the lightcone coordinates of these points, with the Poisson structure obtained from the Peierls bracket. In AdS$_2$ with imaginary radius squared, quantization of the string's mass-squared in momentum-fraction space yields a  P\" oschl--Teller equation, reproducing the SYK operator spectrum.

\end{abstract}

\maketitle

\noindent \textbf{1. Introduction.}
The SYK model has emerged as a prominent example of a strongly interacting quantum system that is both solvable in the large-$N$ limit and closely connected to quantum gravity in AdS$_2$ \cite{1993sachdev, kitaev, Polchinski:2016xgd, Maldacena:2016hyu}. Originally introduced by Sachdev and Ye as a spin model~\cite{1993sachdev}, it was later reformulated by Kitaev~\cite{kitaev} in terms of $N$ Majorana fermions with random all-to-all interactions among $\psyk$ fermions at a time,
\be
\label{eq:sykham}
  H = i^{\psyk/2} \sum_{1 \leq i_1 < i_2 < \ldots < i_\psyk \leq N}
  J_{i_1\ldots i_\psyk} \, \psi_{i_1} \cdots \psi_{i_\psyk} \, .
\ee
The couplings $J_{i_1\cdots i_\psyk}$ are independent Gaussian random variables with zero mean and variance $\langle J_{i_1\cdots i_\psyk}^2 \rangle = \frac{J^2(\psyk-1)!}{N^{\psyk-1}}$.
In the infrared (IR) limit at large $N$, the model exhibits an emergent time-reparametrization symmetry, spontaneously broken by the vacuum to $SL(2)$. The Schwinger--Dyson equations are solved by the dressed two-point function
\begin{equation}
  G(\tau) =  b\, \frac{\mathrm{sgn}(\tau)}{|J\tau|^{2\Delta}},
\end{equation}
with conformal dimension $\Delta = 1/\psyk$ and a constant $b = b(\Delta)$~\cite{kitaev, Polchinski:2016xgd, Maldacena:2016hyu}. The associated Goldstone modes are reparametrization modes whose effective action reduces to the Schwarzian~\cite{Maldacena:2016hyu, Jensen:2016pah, Maldacena:2016upp} due to explicit symmetry breaking.
Beyond two-point functions, the four-point function has been computed in the large-$N$ limit,
\bea
 && \frac{1}{N^2}\sum_{i,j}^N \Bigl\langle T( \psi_i(\tau_1)\psi_i(\tau_2)\psi_j(\tau_3)\psi_j(\tau_4) ) \Bigr\rangle  \\
  \nonumber
 && \qquad  =  G(\tau_{12}) G(\tau_{34}) + \frac{1}{N} \mathcal{F}(\tau_1, \tau_2,\tau_3, \tau_4) + \mathcal{O}(N^{-2}) \, .
\eea
\begin{figure}[h]
\begin{center}
\includegraphics[width=9.0cm]{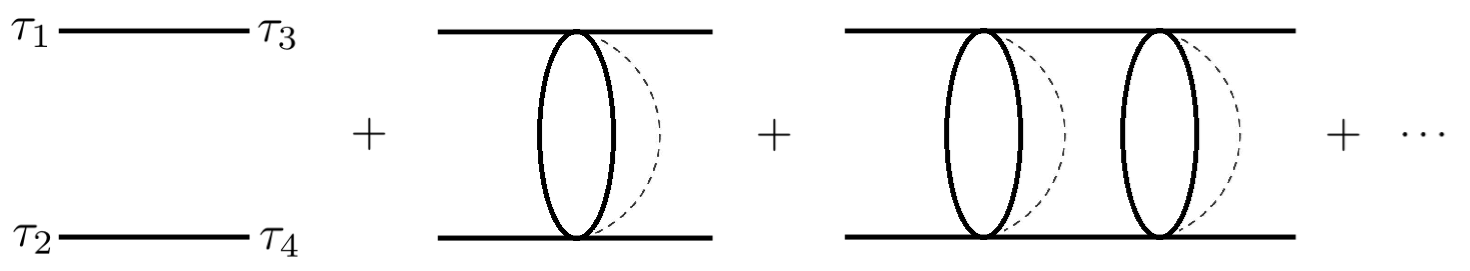}
\caption{\label{fig:ladder}
Ladder diagrams generated by repeated multiplication with the kernel (a vertical ladder rung).
The disorder average links indices at vertices, indicated by dotted lines.
Diagrams with $\tau_3 \leftrightarrow \tau_4$ are not shown. 
}
\end{center}
\end{figure}
The leading connected contribution is contained in $\mathcal{F}$.
This term contains an infinite series of ladder diagrams, as in Figure~\ref{fig:ladder}.
The disconnected contribution is
\[
\mathcal{F}_0(\tau_1, \tau_2, \tau_3, \tau_4) = -G(\tau_{13}) G(\tau_{24}) + G(\tau_{14}) G(\tau_{23}) \, .
\]
A ladder diagram with $n{+}1$ rungs can be obtained from the $n$-rung diagram via the 
integral operator
\be
\label{eq:graphb}
\mathcal{F}_{n+1}(\tau_1, \ldots, \tau_4) =
\int d\tau \, d\tau'  K(\tau_1, \tau_2; \tau, \tau') \, \mathcal{F}_n(\tau, \tau', \tau_3, \tau_4) \, ,
\ee
where the kernel is
\[
K(\tau_1, \tau_2; \tau_3, \tau_4) \equiv -J^2 (\psyk - 1) \, G(\tau_{13}) G(\tau_{24}) \, G(\tau_{34})^{\psyk - 2} \, .
\]
The result can be expressed in an operator product expansion (OPE) form,
\[
\mathcal{F} = \sum_{n=0}^\infty \mathcal{F}_n \ \sim \ \sum_{m=1}^{\infty} c_m^2 \, \chi^{h_m} \, {}_2F_1(h_m, h_m, 2h_m; \chi) \, ,
\]
where $\chi = \tfrac{\tau_{12}\tau_{34}}{\tau_{13}\tau_{24}}$ is the cross-ratio (with $\tau_{ij} \equiv \tau_i - \tau_j$), $h_m$ are the conformal dimensions of the exchanged operators, and $c_m$ are the corresponding OPE coefficients.
The symbol ``$\sim$'' indicates that we omit the infinite contribution from the $h_0 = 2$ gravity mode, which must be treated away from the conformal limit.

The kernel can be diagonalized by exploiting $SL(2)$ invariance.
The spectrum of exchanged bilinear operators in the 4-point function is determined by $k(h) = 1$, where
\be
\label{eq:kh}
  k(h) \equiv   -(\psyk - 1)
\frac{
\Gamma(\tfrac{3}{2} - \tfrac{1}{\psyk})
\Gamma(1 - \tfrac{1}{\psyk})
\Gamma( \tfrac{1}{\psyk} + \tfrac{h}{2} )
\Gamma( \tfrac{1}{2} + \tfrac{1}{\psyk} - \tfrac{h}{2} )
}{
\Gamma(\tfrac{1}{2} + \tfrac{1}{\psyk})
\Gamma(\tfrac{1}{\psyk})
\Gamma( \tfrac{3}{2} - \tfrac{1}{\psyk} - \tfrac{h}{2} )
\Gamma( 1 - \tfrac{1}{\psyk} + \tfrac{h}{2} )
}\, .
\ee
For $\psyk=4$ one finds $h = 2, \ 3.77, \ 5.68, \ 7.63, \ 9.60, \ldots$~\cite{Maldacena:2016hyu}.
In this Letter, we reproduce this spectrum (for any $\psyk$) from the mass spectrum of a folded string in rigid AdS$_2$ with ``imaginary radius-squared.'' To develop the necessary tools, we begin with the string in flat space and standard AdS$_2$.

\begin{figure}[h]
\begin{center}
\raisebox{0.4cm}{\includegraphics[height=2.65cm]{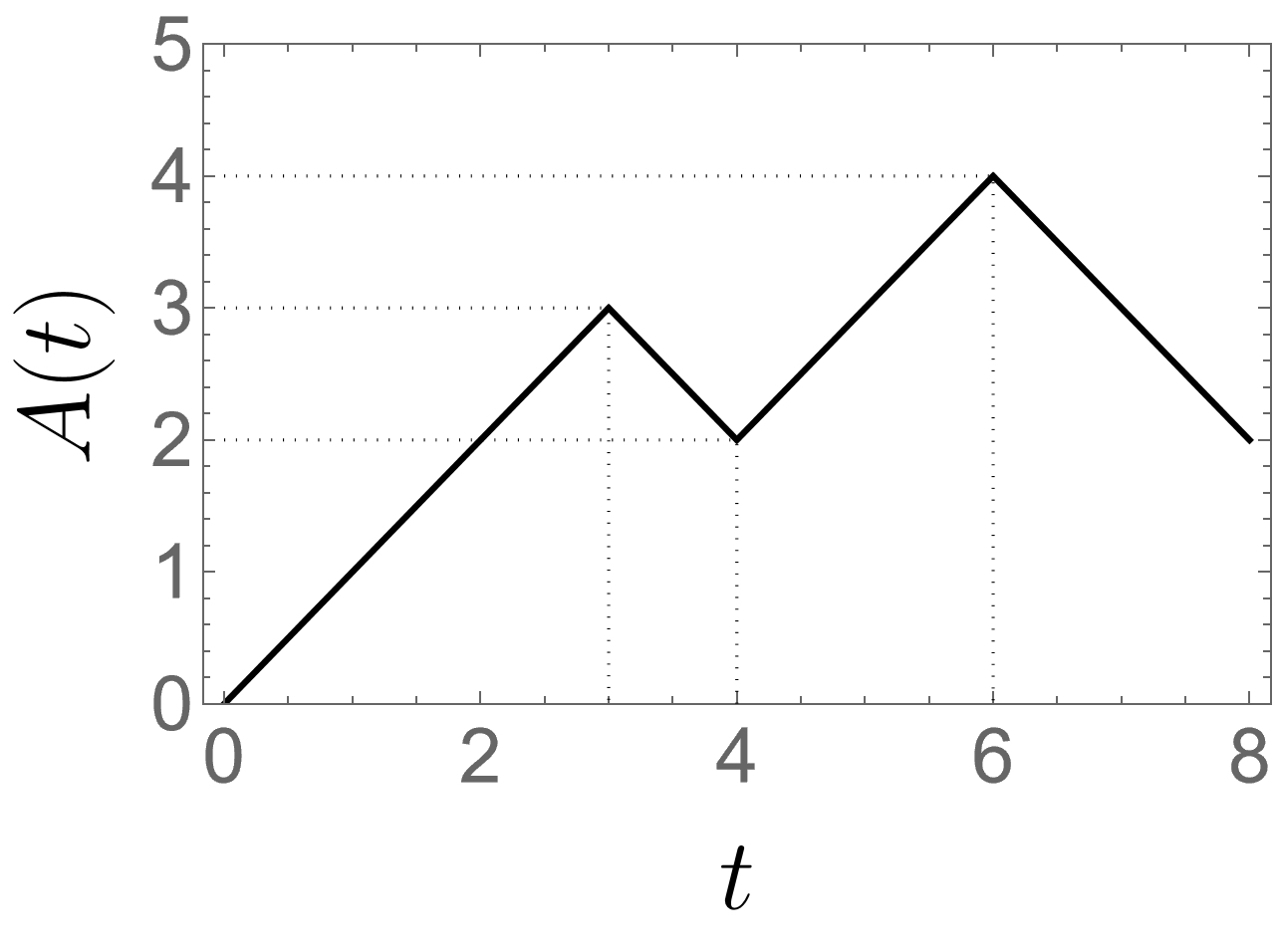} }\quad
\includegraphics[height=4.25cm]{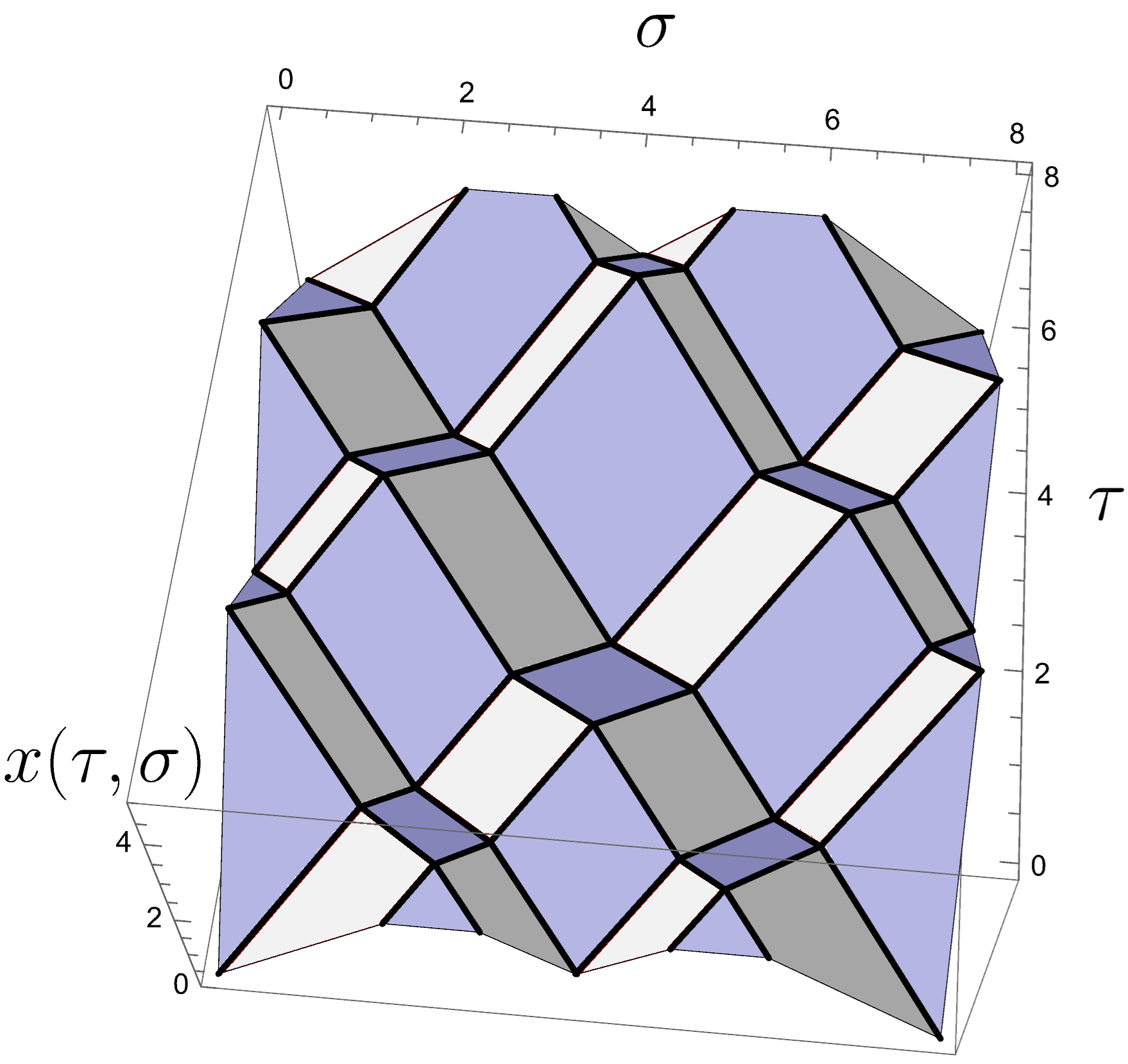}
\caption{\label{fig:afun}
{\it Left:} The directrix function. {\it Right:} 3d plot of the embedding function $x(\tau,\sigma)$. Dark and light blue regions map into left- and right-moving particle worldlines. White/gray patches map into strings stretching between the particles.
}
\end{center}
\end{figure}

\begin{figure}[h]
\begin{center}
\includegraphics[width=8cm]{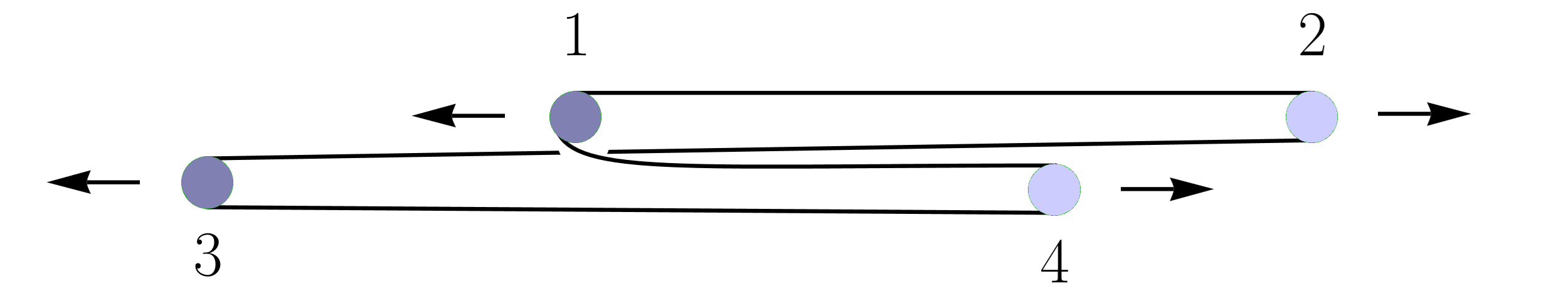}
\caption{\label{fig:fourfolded} Closed folded string with four particles ($L=4$).
}
\end{center}
\end{figure}

\bigskip
 
\noindent \textbf{2. Folded strings in $\RR^{1,1}$.}
We parametrize the string worldsheet by $(\tau, \sigma)$ and the target space by $(t,x)$.
In temporal conformal gauge, $t(\tau, \sigma) = \tau$, and the string embedding is given by $x = x(\tau,\sigma)$.
The equation of motion is the free wave equation 
\[
 \p_\tau^2 x - \p_\sigma^2 x = 0 \, ,
\]
supplemented by the constraints $\p_\tau x \, \p_\sigma x = 0$ and $(\p_\tau x)^2 + (\p_\sigma x)^2 = 1$.
Open string solutions can be constructed using a {\it directrix} function $\mathcal{A}(\xi) $, which describes the motion of an endpoint of the string \cite{Soederberg},
\be
  \nonumber
  x(\tau,\sigma) =  \half\le[ \mathcal{A}(\tau+\sigma)+\mathcal{A}(\tau-\sigma)\ri] \, .
\ee
The directrix satisfies $|\mathcal{A}'| = 1$ and the periodicity condition
$\mathcal{A}(\xi) + 2P = \mathcal{A}(\xi+2E)$, where $P$ and $E$ are the total momentum and energy, respectively.

An example of a directrix is shown in Figure~\ref{fig:afun} (left) with $E = 4$ and $P = 1$.
The corresponding piecewise-linear embedding $x(\tau, \sigma)$ is shown in the right panel as a 3d plot.
Dark and light blue regions correspond to $\partial_\tau x = \pm 1$ and $\partial_\sigma x = 0$, and are mapped into null worldlines of particles in target space.
White and gray regions, with $\partial_\sigma x = \pm 1$ and $\partial_\tau x = 0$, are mapped into string segments stretched between the particles. Henceforth, we will denote the number of particles on the string by $L$. A closed string example with $L=4$ is depicted in Figure~\ref{fig:fourfolded}.

\bigskip

\begin{figure}[h]
\begin{center}
\includegraphics[height=4.5cm]{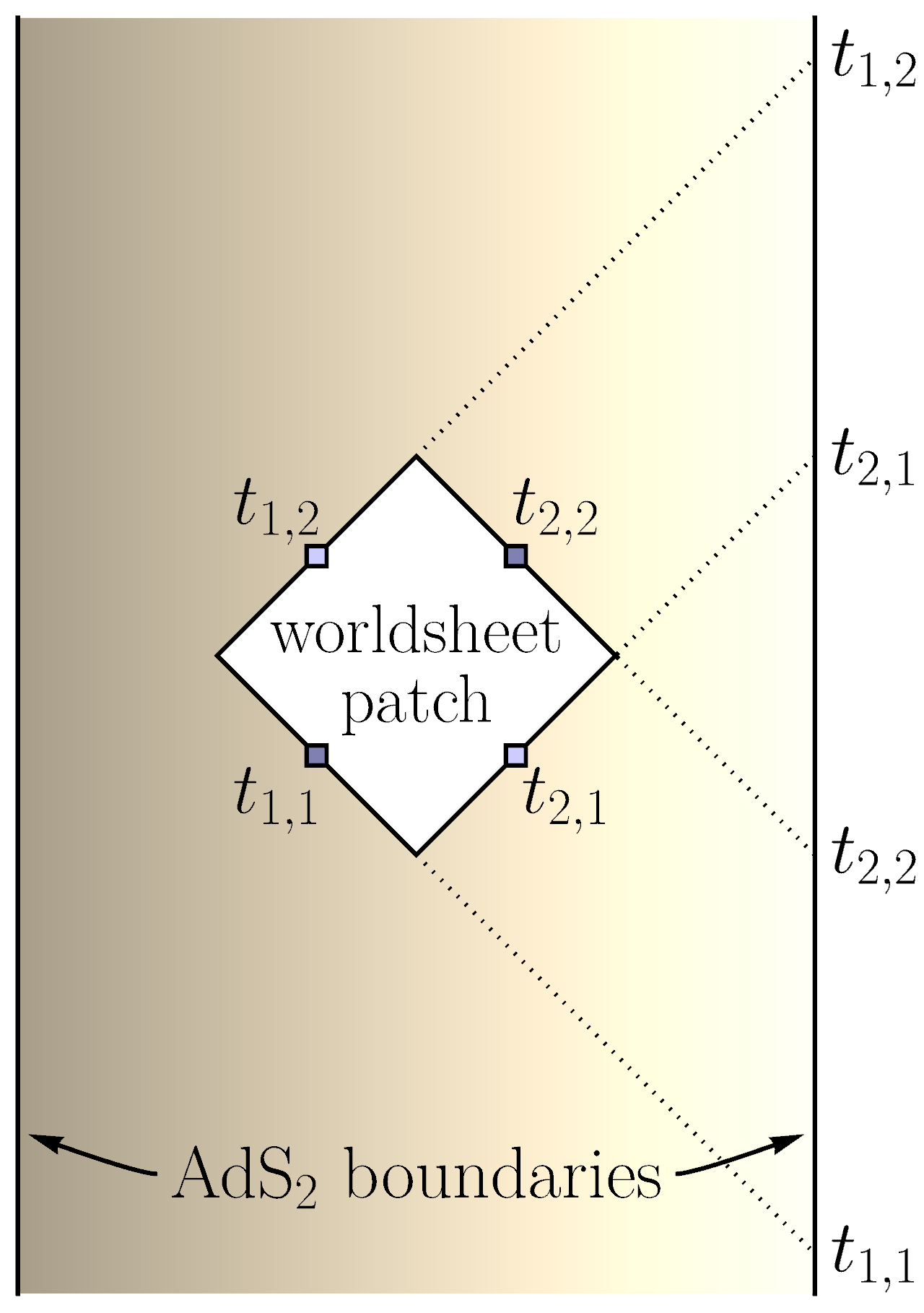}\quad
\raisebox{0.4cm}{\includegraphics[width=4.8cm]{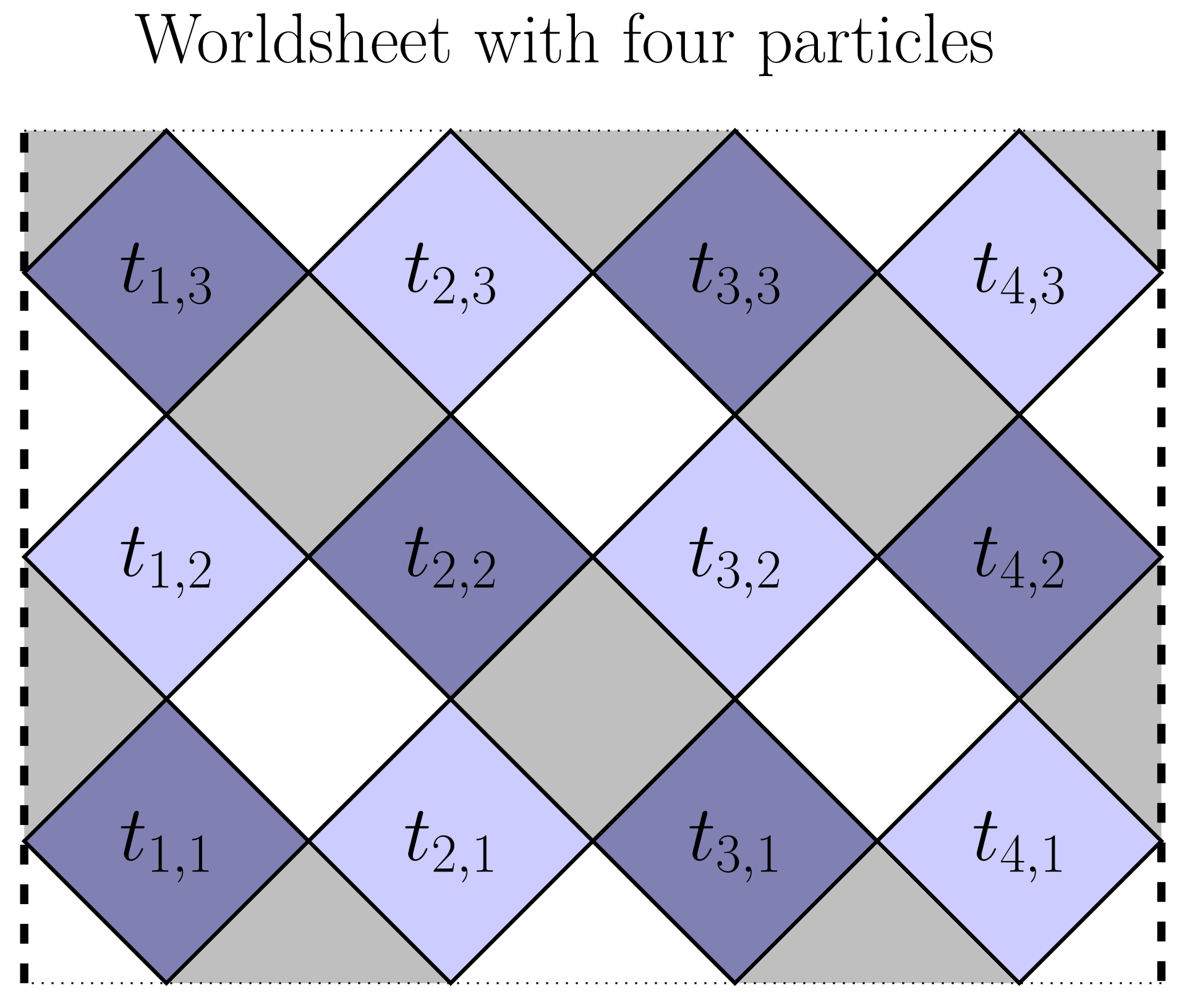}}
\caption{\label{fig:super}
{\it Left:} Boundary time coordinates $t_{i,j}$ for a worldsheet patch. {\it Right:}
The worldsheet of a folded string in AdS$_2$. Dashed lines are identified. The string has four particles ($L=4$) connected by four string segments in a chain.
Dark and light blue patches correspond to left- and right-moving particles.
Gray and white patches are mapped into string segments stretching between the particles.
 }
\end{center}
\end{figure}

\begin{figure}[h]
\begin{center}
\includegraphics[width=6cm]{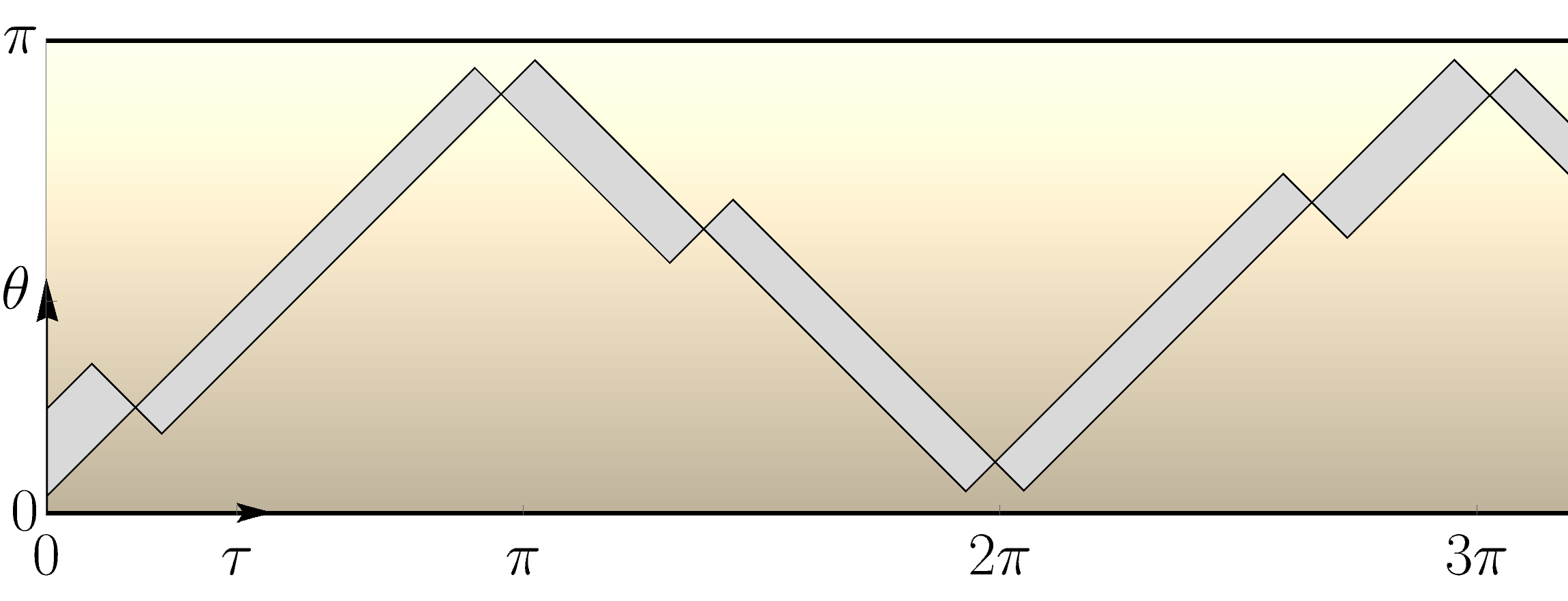} \hskip 0.5cm
\raisebox{0.2cm}{\includegraphics[width=2.0cm]{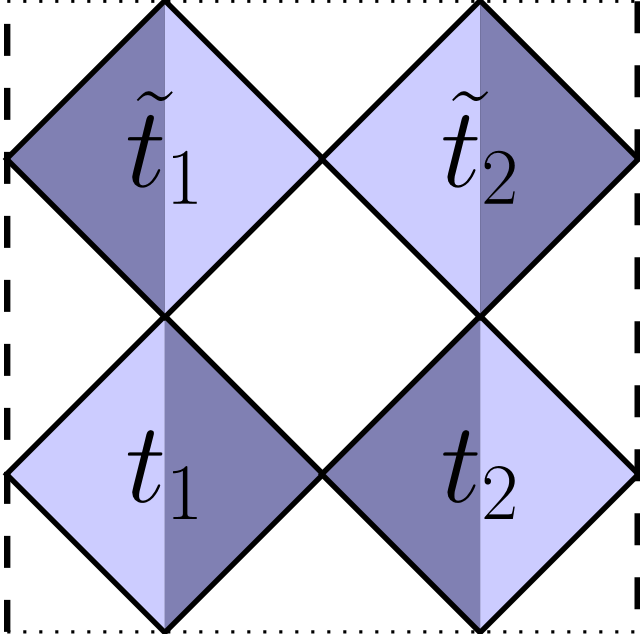}}
\caption{\label{fig:globalmotion}
{\it Left:} Folded string with two particles ($L=2$) in AdS$_2$. Global time $\tau$ runs to the right; horizontal lines at $\theta=0, \pi$ denote the boundaries.  {\it Right:} Phase space consists of two rows of time variables: $t_i \equiv t_{i,j}$, $\tilde{t}_i \equiv t_{i,j+1}$ for a fixed $j$ ($i=1,2$). Whether a blue patch maps to a left- or right-moving particle depends on $j \bmod 2$, hence the split coloring.}
\end{center}
\end{figure}

\noindent \textbf{3. Folded strings in AdS$_2$.} The setup has been discussed in~\cite{Bars:1994sv, Bars:1994xi} (along with generalizations to arbitrary target space metrics). 
The worldsheet is depicted in Figure~\ref{fig:super} (right) and is structurally analogous to the flat-space case of Figure~\ref{fig:afun} (right). 
Dark (light) blue regions are mapped into massless particles moving to the left (right).
The embedding is fully specified by the fixed lightcone coordinates of these worldlines, see Figure~\ref{fig:super}, left. We will use ``boundary time variables'' $t_{i,j}$ defined for left-moving particles as $t_{i,j} = t-z$, and for right-moving particles as $t_{i,j} = t+z$, where $t,z$ are Poincar\'e coordinates of the particle, and the metric is given by
\[
  ds^2 = R^2 {-dt^2 + dz^2 \ov z^2  } \, ,
\]
where $R$ is the AdS$_2$ radius.
The area of the patch in Figure~\ref{fig:super} (left) can be expressed in terms of the surrounding boundary time variables~\cite{Vegh:2016hwq, Vegh:2016fcm}:
\be
  \nonumber
  A = 2 R^2\log\le| {(t_{1,2}-t_{2,2})(t_{1,1}-t_{2,1}) \ov (t_{1,2}-t_{1,1})(t_{2,2}-t_{2,1}) }\ri| \, .
\ee
The Nambu--Goto action is the total induced worldsheet area multiplied by $-\frac{1}{2\pi \alpha'}$, where $\alpha'$ is the square of the string length.
It can be expressed as 
\be
  \label{eq:action}
S[t] = -4g  \sum_{j} \sum_{i=1}^L\left[
     \log|t_{i,j} - t_{i+1,j}|-\log|t_{i,j} - t_{i,j+1}|
\right] \, ,
\ee
where $g \equiv  {R^2 \ov 2\pi \alpha'}$,  and $L$ is the number of particles on the string. 
The variables $t_{i,j}$ are real and for a closed string we have $t_{i+L, j} = t_{i, j}$.
The equation of motion reads
\be
  \label{eq:eom}
  \hskip -0.1cm {1\ov t_{i,j} - t_{i,j+1}}+   {1\ov t_{i,j} - t_{i,j-1}} =
  {1\ov t_{i,j} - t_{i+1,j}}+   {1\ov t_{i,j} - t_{i-1,j}} \, ,
\ee
which couples each site to its four nearest neighbors.
Cauchy data are provided by specifying two adjacent rows of $t_{i,j}$ variables. An example solution with $L=2$ particles is shown in Figure \ref{fig:globalmotion} (left).

\bigskip

\noindent \textbf{4. The Peierls bracket.} We now turn to the Hamiltonian formalism, parametrizing phase space by $t_{i,j}$. Their Poisson brackets are computed using Peierls' construction~\cite{Peierls:1952cb}, applicable to relativistic theories.
For a general two-derivative Lagrangian field theory with action
\[
S = \int d^d x \, \sqrt{-g} \, \mathcal{L}(\phi, \partial\phi, \partial^2\phi, \ldots)
\]
the Peierls bracket of functionals $f[\phi]$, $h[\phi]$ is defined as
\be
\label{eq:pei}
\{f, h\}[\phi] \equiv \left. \frac{d}{d\lambda} f[\phi + \lambda \, \delta_h \phi] \right|_{\lambda=0}
\ee
where $\phi$ is any solution to the equations of motion.
To construct $\delta_h \phi$, we deform the action as $S' = S - \lambda h$
and solve the modified equations of motion to first order in $\lambda$ for two configurations:
\[
\phi + \lambda\, \delta\phi^R \quad \text{(retarded)}, \qquad \phi + \lambda\, \delta\phi^A \quad \text{(advanced)},
\]
where the retarded (advanced) solution vanishes at sufficiently early (late) times.
These are well-defined if $f$ and $h$ have support only in a finite time interval.
Then, we define $\delta_h \phi \equiv \delta\phi^R - \delta\phi^A$.

\begin{figure}[h]
\begin{center}
\includegraphics[width=2.5cm]{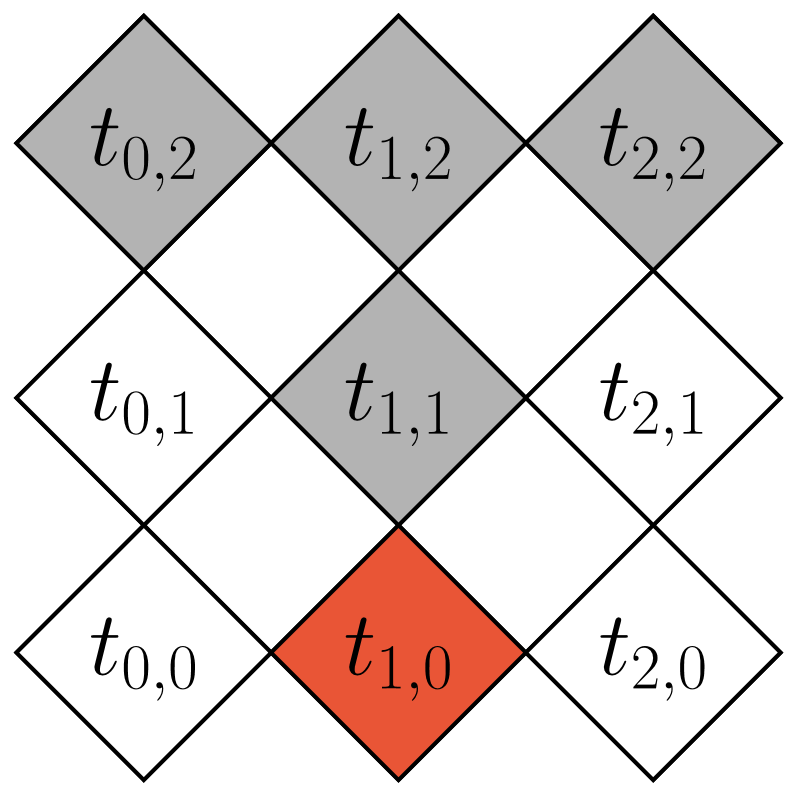}
\caption{\label{fig:repsonse}
Turning on a source for $h=t_{1,0}$.
Sites inside the discrete forward lightcone of $(1,0)$ are indicated by gray color. The retarded perturbation $ \delta\phi^R $ is non-zero in this region.
 }
\end{center}
\end{figure}

Although the folded string action is defined on a discrete spacetime lattice, the solutions are embedded into the continuum Nambu--Goto theory, so the Peierls bracket can still be computed.
Here, the role of the field $\phi$ is played by $t_{i,j}$, and the perturbations $\delta\phi^{R,A}_{i,j}$ are likewise discretized.
Figure~\ref{fig:repsonse} shows the support of the retarded perturbation when $h$ is chosen to be $t_{1,0}$.
As discussed in Appendix A, \eqref{eq:pei} yields the Poisson bracket
\be
  \label{eq:poissonb2}
  \{ t_{i,j+1}, \, t_{i,j} \}  =  {1\ov 4g} (t_{i,j+1}-t_{i,j})^2 \, ,
\ee
together with $\{ t_{i+1,j}, \, t_{i,j} \}  = 0$. 
These brackets were previously considered in \cite{GEKHTMAN2016390}. 
The phase space is parametrized by the $2L$ time variables on two consecutive slices. Henceforth we use the notation
\be
  \label{eq:deft}
  t_i \equiv t_{i,j} \quad \text{and} \quad \tilde{t}_i \equiv t_{i,j+1} \, ,
\ee
with $t_{i+L} = t_{i}$ and $\tilde{t}_{i+L} = \tilde{t}_{i}$ for a closed string.
As only the two-particle case is relevant for SYK, we will fix $L=2$ for the remainder of the paper.

\bigskip

\begin{figure}[h]
\begin{center}
\includegraphics[height=2.85cm]{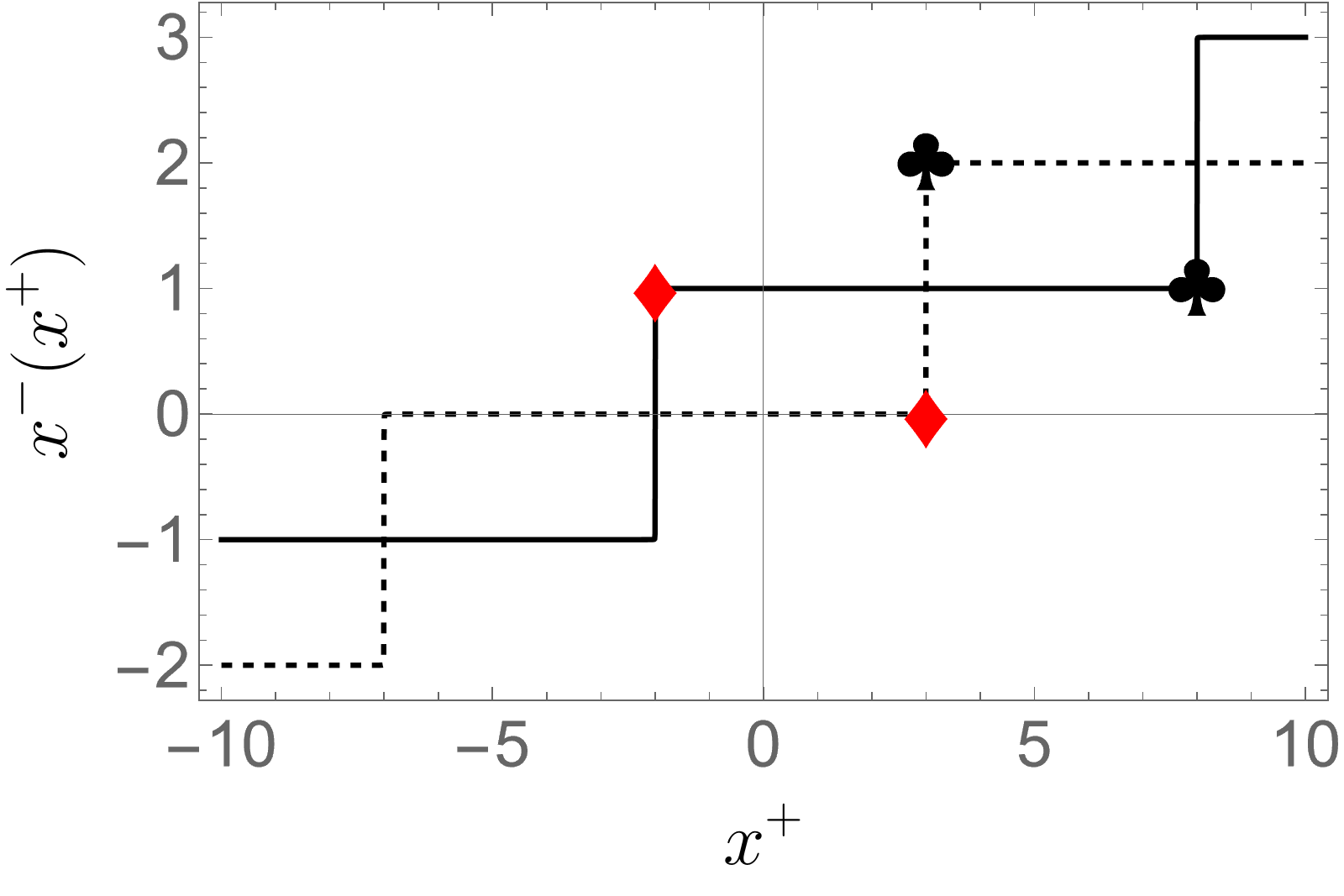}
\includegraphics[height=2.85cm]{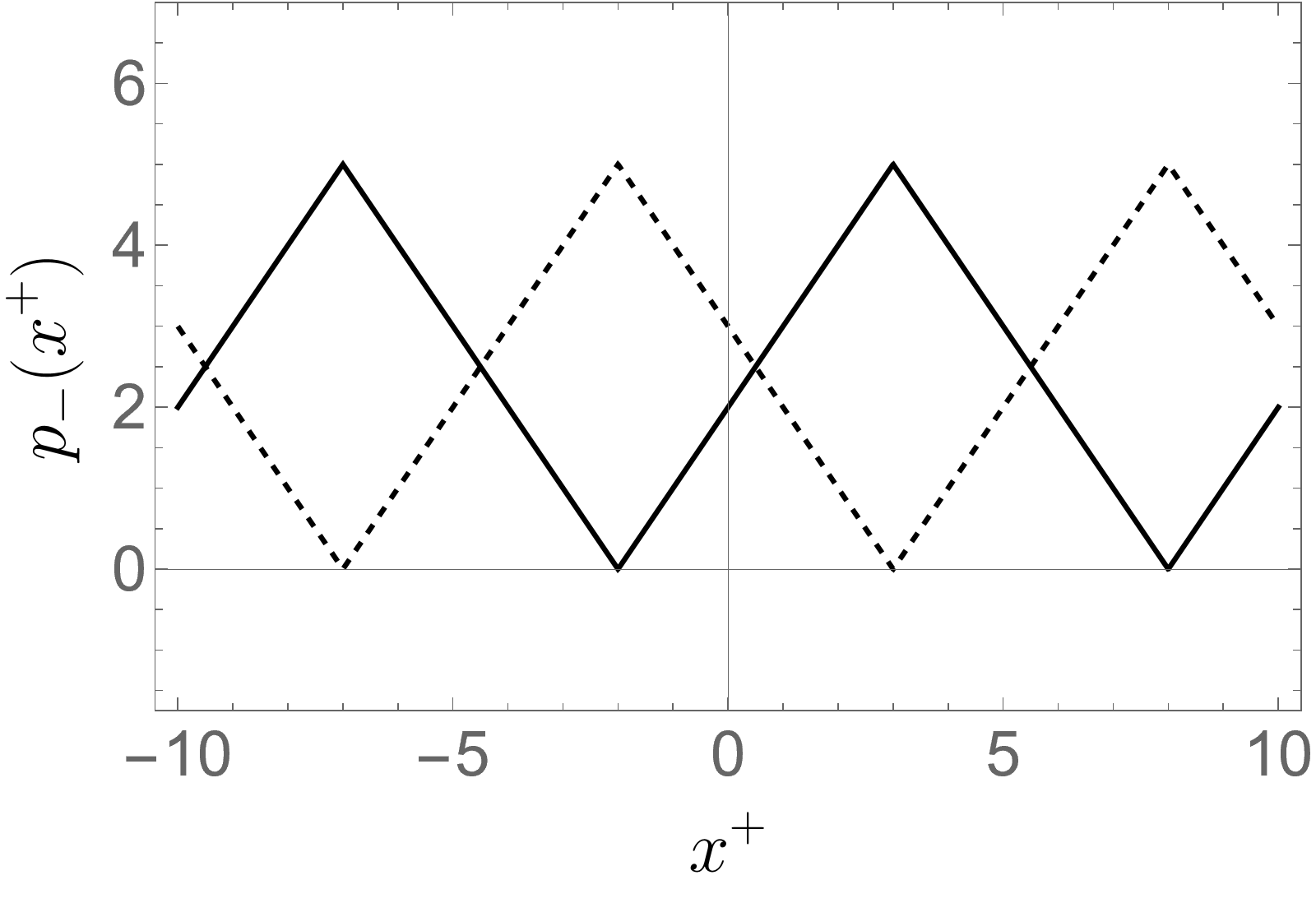}
\caption{\label{fig:numeric}
{\it Left:} Numerically calculated particle trajectories using Hamiltonian \eqref{eq:lcone1} with negligible masses ($m_{1,2}=10^{-2}$). Diamonds and clubs mark turning points referred to in the main text. {\it Right:} The conjugate momenta remain positive.
}
\end{center}
\end{figure}

\noindent \textbf{5. The Hamiltonian in flat space.}
To gain some familiarity, we first discuss the case of a flat target space.
In lightcone frame, the system of two massive particles connected by a linear potential has the Hamiltonian \cite{Lenz:1995tj}
\be
  \label{eq:lcone1}
  H^-(x^-_i, p^+_{i}) = {m_1^2 \ov 2 p^+_1(x^+)}+{m_2^2 \ov 2 p^+_2(x^+)} + \kappa|x^-_1(x^+)-x^-_2(x^+)| \, ,
\ee
where $x^-_i= {t_i - x_i \ov \sqrt{2}}$  are the lightcone positions of the particles, $p^+_i$ are the conjugate momenta, $x^+= {t + x \ov \sqrt{2}}$  is the time variable, and $\kappa$ is the string tension.
To describe a folded string, the massless limit will be of interest. Note that $m_i$ cannot simply be set to zero, because the kinetic terms are important whenever the momenta vanish. For negligible masses, Hamilton's equations dictate that  $x^-_i$ remain constant almost everywhere and that $p^+_i$ evolve linearly in time, bouncing back when they reach zero. At the same time, the corresponding particle position undergoes a jump, as shown in Figure \ref{fig:numeric}.

$H^-$, together with $H^+  =   p^+_1 +  p^+_2$ and $K =  x^-_1 p^+_1 + x^-_2 p^+_2 $, 
generates the \poincare algebra 
\[
  \{H^+, H^- \} = 0 \, , \ \
  \{K, H^- \} = -H^- \, , \ \
  \{K, H^+ \} = H^+ \, .
\]
The mass-shell condition reads
\be
  \label{eq:flatmasss}
  m^2 = 2H^+ H^- = {m_1^2 \ov z}+{m_2^2 \ov 1-z}
  + 2\kappa |s| \, ,
\ee
where we have introduced the momentum fraction variable $z$ and its conjugate $s$  
\be
  \label{eq:zsdef}
  z \equiv { p^+_1 \ov  p^+_1 +  p^+_2} \, , \qquad s \equiv (x^-_1-x^-_2)(p^+_1 +  p^+_2 ) \, ,
\ee
with $ \{ s, z \} = 1$.

At the quantum level, $m^2$ commutes with all operators and thus shares eigenfunctions with the Hamiltonian. Rather than quantizing \eqref{eq:lcone1}, which yields a 2d \schr equation, one can quantize \eqref{eq:flatmasss}, leading to the somewhat simpler 1d ’t~Hooft equation \cite{tHooft:1974pnl}. This derivation requires projecting the wavefunction onto positive-momentum states, so that $\psi(z)$ has support only on $z \in [0,1]$ (see \cite{Lenz:1995tj}). The projection is also motivated by the time evolution of the classical momenta, which stay positive; see Figure~\ref{fig:numeric} (right).

We now want to generalize the mass-shell condition \eqref{eq:flatmasss} to AdS$_2$, but immediately encounter a problem: the analog of $H^-$ involves explicit $x^+$ dependence in the two-particle potential $V^- \propto |\cot(x^+ - x^-_1)-\cot(x^+ - x^-_2) |$, so $m^2$ cannot be written in the same simple form as in flat space.
 
Instead, let us return to flat space and set $m_i=0$. In the center-of-mass frame, the particles reverse direction simultaneously, and the total mass is determined by the maximal extent of the string,
\be
  \label{eq:flatmasss2}
  m^2 =  2\kappa^2(x^-_1-x^-_2)(x^+_2-x^+_1) \, ,
\ee
where we assume that the turning points $(x^+_1, x^-_1)$ and $(x^+_2, x^-_2)$ are spacelike separated. 
Equation~\eqref{eq:flatmasss2} depends only on the lightcone coordinates $x_i^\pm$, which will be used to parametrize phase space.

At a given time, points in phase space correspond to classical solutions of the equations of motion, so those representing the same solution must be identified. Thus, {\it phase space points related by discrete time steps of the turning points must be identified}.
For example, in Figure~\ref{fig:numeric}, turning points marked by diamonds and clubs correspond to 
the same classical solution. Thus, as points in phase space, the diamonds are equivalent to the clubs.

\bigskip
 
\noindent \textbf{6. The Hamiltonian in AdS$_2$.} 
Appendix B derives the string's mass-squared in AdS$_2$ using the center-of-mass Hamiltonian in a similar way. In terms of the time variables $t_{i,j}$ introduced in Section 3, the result is
\be
  \label{eq:casi1}
  M^2 \equiv (mR)^2 = 16g^2{ (\tilde t_1 - t_2)(\tilde t_2 - t_1) \ov (t_1 - \tilde t_1)(\tilde t_2 - t_2)} \, ,
\ee
where $g \equiv  {R^2 \ov 2\pi \alpha'}$ and we used the notation \eqref{eq:deft}. The formula gives the mass-squared on the 4d phase space parametrized by $(t_1, t_2, \tilde t_1, \tilde t_2)$; see also Figure \ref{fig:globalmotion} (right). In a flat space limit, the expression reduces to \eqref{eq:flatmasss2}.  

It is convenient to introduce canonical coordinates \cite{suris},
\be
  \label{eq:xidef}
  q_{i} \equiv {  \tilde t_{i}+ t_{i} \ov 2 } \, , \qquad
  p_{i} \equiv {4g \ov \tilde t_{i} - t_{i}} \, \qquad (i=1,2) \, ,
\ee
which satisfy $ \{ q_{i} ,  p_j \} =  \delta_{ij}$. Let us define the generators  
\be
  \label{eq:sl2gen}
  S^+_i \equiv  p_i \, , \qquad
  S^0_i \equiv q_i p_i \, , \qquad
  S^-_i \equiv  -t_i \tilde t_i  p_i \, ,
\ee
which obey the $SL(2)\times SL(2)$ algebra  
\be
  \label{eq:algebra}
  \{ S^0_i, S^\pm_j \} = \pm S^\pm_i \delta_{ij} \, , \quad
  \{ S^+_i, S^-_j \} = 2 S^0_i \delta_{ij}\, .
\ee
The expressions in \eqref{eq:sl2gen} can be motivated by their appearance as matrix elements of the $2\times 2$ Lax matrix associated to the system (see Section 4.1 in \cite{Vegh:2023snc}).
Note that defining the total spin as $\vec S = \vec S_1 + \vec S_2$, the RHS of eqn.~\eqref{eq:casi1} can be rewritten using the quadratic Casimir
\be
  \label{eq:casi1}
  M^2 =   - (S^0)^2 - \half(S^+ S^- + S^- S^+)\, .
\ee
Finally, we express $M^2$ in terms of a momentum fraction variable $z$, as in the flat-space case \eqref{eq:flatmasss}.
For an unfixed pair $(y_1, y_2)$, we define $(z,s, P_+, Q^+)$ as follows:
\be
  \label{eq:cootransf}
  P_+ \equiv p_1+p_2 \, , \
  z \equiv {p_1 \ov P_+} \, , \
  s \equiv ( y_1 - y_2) P_+ \, , \
  Q^+ \equiv y_2 + {z s\ov P_+} \, .
\ee
Here $z$ and $s$ are analogous to \eqref{eq:zsdef}. Note that in \eqref{eq:zsdef}, the definition of $s$ involves the $x^-$ variables, which we aim to generalize to AdS$_2$.
From \eqref{eq:casi1} we find  
\be
\label{eq:twocas}
  M^2 =
\begin{cases}
  \  sz(1-z)s + 4gs \quad  & \textrm{for} \ (y_1, y_2) = (t_1, \tilde t_2) \, , \\
  \  sz(1-z)s - 4gs \quad & \textrm{for}  \ (y_1, y_2) = (\tilde t_1,  t_2) \, .
\end{cases}
\ee
In both cases, $\{ s, z \} = 1$ and $\{ Q^+, P_+ \} = 1$, with all the other brackets vanishing.
As the string oscillates, at every half-period $z$ reaches $0$ or $1$, and the particles collide. At this point, we perform a discrete forward time step (Figure \ref{fig:super}, right), replacing $t_i \equiv t_{i,j}$ and $ \tilde t_i \equiv t_{i,j+1}$ with $t_{i,j+1}$ and $t_{i,j+2}$, respectively. Since $y_i$ should be identified with the right-moving ($x^-$-type) time variables, $M^2$ also switches to the other expression in \eqref{eq:twocas} at each half-period.
Because $\tilde t_i > t_i$ and $P_+ > 0$, $s$ changes sign at these steps. Thus, we can combine \eqref{eq:twocas} into
\be
  \label{eq:absvalue}
  M^2 = sz(1-z)s + 4g|s| \, ,
\ee
which is the AdS$_2$ generalization of the classical ’t~Hooft equation for massless particles. We note that an explicit map to the center-of-mass frame exists \cite{Vegh:2023snc}, under which $M$ is mapped to the center-of-mass Hamiltonian.  

\bigskip

\noindent \textbf{7. Folded strings in ``complex AdS$_2$'' and SYK.}  
For applications to SYK, we will set $g$ to the value
\be
\label{eq:setg}
g_0 = {i \ov 4}(1-2 \Delta)  \, ,   
\ee
where $\Delta = 1/\psyk$ is the IR conformal dimension in the theory with $\psyk$-fermion interactions, see \eqref{eq:sykham}.  We motivate this value by the following observation.
The fields $t_{i,j}$ transform under the global $\mathrm{SL}(2,\mathbb{R})$ symmetry via M\"obius transformations  $t \mapsto \tfrac{at + b}{ct + d}$ where $ad - bc = 1$.
By choosing an $SL(2)$-invariant measure (see also the lattice Schwarzian theory in \cite{Stanford:2017thb}), we can write the path integral  
\bea
\label{eq:path}
&& Z =  \int \prod_{i,j}   \frac{dt_{i,j} }{|t_{i,j} - t_{i,j+1}|} \times \\
&&
\nonumber
\times e^{ -4ig \sum_{j} \sum_{i=1}^L
\left\{
    \log|t_{i,j} - t_{i+1,j}|-\log|t_{i,j} - t_{i,j+1}|
\right\} }.
\eea
If we set $g=-g_0$, then after bringing the arguments of the logarithms down from the exponent, eqn.~\eqref{eq:path} takes precisely the form of a coordinate-space  Feynman diagram.
Notably, for the simplest folded string with two particles ($L=2$) this Lorentzian path integral matches the Euclidean SYK ladder diagrams in Figure \ref{fig:ladder} (up to signs and constant factors in the fermion propagators). 

Returning to the Hamiltonian formalism, we now set $g=g_0$ to match the conventions in \cite{Vegh:2024uie} and attempt to make sense of the system. (Switching the sign corresponds to taking the shadow conformal dimension $\Delta \to 1-\Delta$, which gives the same spectrum). Simply plugging an imaginary $g$ into either formula in \eqref{eq:twocas} produces a complex equation. Instead, we choose symmetric coordinates $y_i \to  {  \tilde t_{i}+ t_{i} \ov 2 }$
in \eqref{eq:cootransf}. Then, the Casimir becomes
\be
  \label{eq:casimag}
  M^2 =  s z(1-z) s +{(\half - \Delta)^2 \ov z(1-z)} \, ,
\ee
and $s$ and $z$ again form a canonical pair.
The repulsive potential keeps $z$ in the middle of the  $[0,1]$ interval, so the two particles never collide (which would occur at $z=0$ or $1$).
Hence, {\it discrete time steps do not need to be performed, and in the absence of absolute values the system simplifies significantly} (cf. eqn. \eqref{eq:absvalue}).
 
\begin{figure}[h]
\begin{center}
\includegraphics[width=2.3cm]{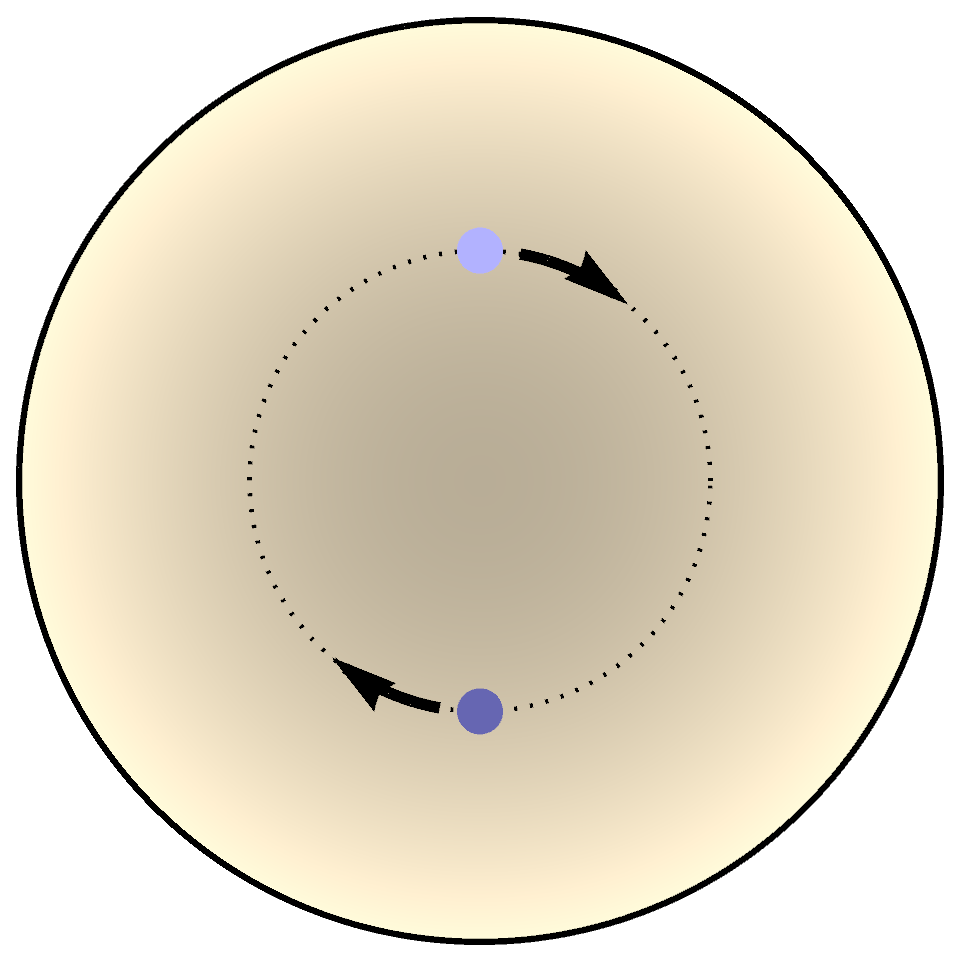}
\caption{
\label{fig:rotat}
Time evolution in phase space for $g = {i \ov 4}(1-2 \Delta)$. Phase space is  parametrized by a pair of complex variables $(t_1,  \tilde t_2)$, here depicted as blue dots on the \poincare disk. They orbit their ``center of mass'' as discussed in Appendix B.}
\end{center}
\end{figure}

Since \eqref{eq:casimag} is real, the variables $z$ and $s$ can also remain real if the time variables are complex conjugates of each other, $t_i = (\tilde t_i)^*$. Classical phase space is parametrized by two complex variables ($t_1,  \tilde t_2$) on the upper-half plane and time evolution is a rigid rotation around their “center of mass,” as illustrated in Figure \ref{fig:rotat}.

We now quantize the system by quantizing the $SL(2)$ generators in \eqref{eq:sl2gen}.
After expressing $t_n = q_n - {2g \ov p_n}$ and $\tilde t_n = q_n + {2g \ov p_n}$, symmetrizing the operators, and setting $s_n = i \p_{p_n} \equiv i \p_{n} $, we obtain the differential operators
\be
  \label{eq:sl2genq}
  S^+_n =  p_n \, , \quad
  S^0_n = i p_n \p_{n} + {i\ov 2} \, , \quad
  S^-_n =  p_n \p_{n}^2 + \p_{n} + {4g^2 \ov p_n}  \, ,
\ee
which act on the momentum-space wavefunction $\psi(p_1, p_2)$.
Their commutators satisfy the $SL(2)$ algebra in \eqref{eq:algebra}, with an extra factor of $i$ on the right-hand side. 
We project the wavefunctions onto positive $p_i$, a projection that commutes with the $SL(2)$ generators.
The Casimir equation \eqref{eq:casi1} can be quantized 
\[
    h(h-1) \psi(p_1, p_2)=-\le[{ S^+ S^- + S^- S^+ \ov 2} +  (S^0)^2 \ri]
\psi(p_1, p_2) 
\]
where $h$ is the conformal dimension and $M^2 =h(h-1) $.
Writing $\psi(p_1, p_2) = \varphi(z, P_+)$, the Casimir acts only on the $z$ coordinate, giving
\be
  \label{eq:ptell}
  h(h-1)\varphi(z) =
  \frac{4g^2 \varphi(z)}{z(1 - z)}
+ (1 - 2z) \varphi'(z) - z(1 - z) \varphi''(z) 
\ee
where the $P_+$ dependence is suppressed.
With an appropriate change of variables \cite{Vegh:2024uie}, this equation is transformed into the \teller equation \cite{Poschl:1933zz}, whose solutions are associated Legendre polynomials.

Near the boundaries $z=0, 1$, the solutions behave as
\be
  \label{eq:behav}
  \varphi(z)\approx  \begin{cases}
     \  c_A z^{\Delta - \half} + c_B z^{\half-\Delta} \quad  & \textrm{for} \ z\to 0 \, , \\
  \ \tilde c_A (1-z)^{\Delta - \half} +\tilde c_B (1-z)^{\half-\Delta} \quad & \textrm{for} \ z\to 1 \, .
\end{cases}
\ee
Imposing a simple symmetric boundary condition 
\be
\label{eq:bc}
{c_A \ov  c_B}  = {\tilde c_B  \ov  \tilde c_A} =  {\Delta \ov 1-\Delta}
\ee
makes the Casimir self-adjoint and yields a discrete spectrum for $h$ \cite{Vegh:2024uie}.
Eigenvalues corresponding to antisymmetric wavefunctions match the SYK operator spectrum $k(h)=1$, where $k(h)$ is given in \eqref{eq:kh}, while symmetric wavefunctions reproduce the bosonic SYK spectrum.
Finally, varying the constant on the RHS of \eqref{eq:bc} reproduces the spectrum of the line of CFTs described in \cite{Gross:2017vhb}.

\bigskip

\noindent \textbf{8. Discussion.} In this Letter, we have presented the ``string dual'' of the SYK model at the conformal fixed point. The string mass is quantized and reproduces the operator spectrum of fermion bilinears, in accordance with the AdS/CFT dictionary. The target space is AdS$_2$ of ``complex radius'' ${R^2 } = {i  \pi \alpha' }(\half- \Delta)$ where $\Delta$ is the conformal dimension of the Majorana fermion in the SYK theory.
In standard AdS$_2$, particles on the string constantly collide
(see Figures~\ref{fig:globalmotion}\, \&  \ref{fig:numeric}).  
On the other hand, for a complex AdS radius, particles never collide and the spectrum can be computed from the \schr equation \eqref{eq:ptell}.
The equation is written in momentum-fraction basis, where the boundary conditions take the simple form \eqref{eq:bc}.
Note that due to the boundary conditions, the two particles do not, in general, lie in unitary irreducible representations of $\widetilde{SL}(2, \mathbb{R})$.
Wigner’s classification applies only to free particles, which occurs for $\Delta = \tfrac{1}{2}$, where the vanishing string tension causes the particles to decouple.
In this limit, the particles furnish discrete series representations \cite{Anninos:2023lin, Pethybridge:2024qci}, and the spectrum follows from their tensor product.

There are several research directions that may be worth exploring. Folded strings can be generalized to configurations with multiple folds, involving $L>2$ particles. With complex AdS radius, the path integral \eqref{eq:path} belongs to the class of (anisotropic) `fishnet' diagrams, which form a completely integrable system thanks to the star--triangle relation~\cite{Zamolodchikov:1980mb, Kazakov:2018qbr, Kazakov:2022dbd, Kazakov:2023nyu}.  The theory is integrable and serves as a bulk dual to anisotropic fishnet CFTs in 1d. Fishchain models have been proposed as bulk duals to isotropic fishnet theories \cite{Gromov:2019aku, Gromov:2019bsj, Gromov:2019jfh}; it would be interesting to generalize them to anisotropic models and compare their descriptions with ours.
Other models, such as the long-range Ising model, defects in CFTs, and complex \cite{Gu:2019jub} or supersymmetric \cite{Fu:2016vas} versions of SYK, also merit further study. Another promising direction is generalization to higher dimensions. However, we note that in $d>1$ the Poisson structure becomes more complicated due to the presence of level-matching-type constraints.

The bulk dual of SYK suffers from the fact that the string scale is of the same order as the AdS scale. To have a string propagating on a large AdS$_2$, the fermion should have conformal dimension $\Delta = \half + i s$, with $s~\gg~1$. In that case, however, the strongly attractive \teller potential will have tachyonic bound states, leading to an instability. Addressing this issue likely requires implementing the boundary equivalent of discrete time steps---a feature absent in the original SYK model.

\bigskip

\noindent {\it Acknowledgments.}  
The author is grateful to D. Anninos, T. Anous, J. Chryssanthacopoulos, A. Georgoudis, F.~Haehl, M. Hanada, M. Preti, and  D. Stanford for interesting discussions and feedback.
The author is partially supported by the STFC Consolidated Grant ST/X00063X/1 ``Amplitudes, strings \& duality.''
The author gratefully acknowledges support from the Simons Center for Geometry and Physics, Stony Brook University at which some of the research for this paper was performed.
The author also thanks the organizers of the ``Black hole physics from strongly coupled thermal dynamics'' workshop at the Simons Center, the ``Fishnet QFTs: Integrability, periods and beyond'' conference at the U. of Southampton, and the ``Quantum Gravity, Holography and Quantum Information'' workshop in Natal, Brazil, where some of the results were presented.
No new data were generated or analyzed during this study.

\bigskip

\begin{center}
\textbf{Appendix A -- Peierls bracket}
\end{center}

In this Appendix, we discuss how to apply Peierls' prescription to derive the Poisson brackets for the time variables, as given in equation \eqref{eq:poissonb2}.
For a closed string, we impose periodicity in the $i$-direction on the lattice; this does not affect the calculation.

To compute the Peierls bracket of a variable $t_{i_0,j_0}$ with any other dynamical variable, we introduce a source term in the action,
\[
S'[t] = S[t] - \lambda \, t_{i_0,j_0},
\]
where $S[t]$ is given in \eqref{eq:action}.
This modifies the equations of motion to
\begin{align*}
   {4g\ov t_{i,j} - t_{i,j+1}}+   & {4g\ov t_{i,j} - t_{i,j-1}} = \\
   & {4g\ov t_{i,j} - t_{i+1,j}}+   {4g\ov t_{i,j} - t_{i-1,j}} + \lambda \, \delta_{i,i_0}\delta_{j,j_0}\, .
\end{align*}
Consider the retarded perturbation $\delta\phi_R$ which vanishes in the past.
The perturbed solution evolves identically to the unperturbed one until $j=j_0$.
At $(i,j)=(i_0,j_0)$, the source creates a “wake,” so $\delta\phi_R$ becomes nonzero inside the discrete future lightcone of the point $(i_0,j_0)$ on the lattice.
This causal structure is illustrated in Figure \ref{fig:repsonse} for the example $(i_0,j_0) = (1,0)$.
The advanced perturbation $\delta\phi_A$ is nonzero inside the analogous discrete past lightcone.

For clarity, we drop the zero subscript from now on.
Next, consider the Poisson bracket $\{t_{i,j}, t_{i,j+1}\}$.
The variable $t_{i,j+1}$ can be computed from $t_{i,j}$, $t_{i-1,j}$, $t_{i+1,j}$, and $t_{i,j-1}$ using the equations of motion.
The unperturbed solution for $t_{i,j+1}$ reads
\bea
\nonumber
 && t_{i,j+1} =
 \\
 \nonumber
 &&
\frac{
\left(t_{i-1,j} - t_{i,j-1}\right) t_{i,j}^2 + \left(t_{i-1,j} \left(t_{i,j-1} - 2 t_{i,j}\right) + t_{i,j}^2\right) t_{i+1,j}
}{
t_{i,j}^2 + t_{i-1,j} \left(t_{i,j-1} - t_{i+1,j}\right) + t_{i,j-1} \left(t_{i+1,j}-2 t_{i,j}\right)
}\, .
\eea
Due to the source, this changes by an amount $\lambda \delta\phi^R_{i,j+1}$,
\bea
\nonumber
 &&\delta\phi^R_{i,j+1} = \\
 \nonumber
 &&\frac{(t_{i-1,j}-t_{i,j})^2 (t_{i,j-1}-t_{i,j})^2 (t_{i,j}-t_{i+1,j})^2}{4 g \left(t_{i,j}^2+t_{i-1,j} (t_{i,j-1}-t_{i+1,j})+t_{i,j-1} (t_{i+1,j}-2 t_{i,j})\right)^2} \, .
\eea
Since our phase space is spanned by two rows of independent time variables $t_{i,j}$ and $t_{i,j+1}$,
we express $t_{i,j-1}$ in the above expression in terms of these variables using the unperturbed equations of motion: 
\bea
\nonumber
 && t_{i,j-1} =
 \\
 \nonumber
 &&
\frac{
\left(t_{i-1,j} - t_{i,j+1}\right) t_{i,j}^2 + \left(t_{i-1,j} \left(t_{i,j+1} - 2 t_{i,j}\right) + t_{i,j}^2\right) t_{i+1,j}
}{
t_{i,j}^2 + t_{i-1,j} \left(t_{i,j+1} - t_{i+1,j}\right) + t_{i,j+1} \left(t_{i+1,j}-2 t_{i,j}\right)
}\, .
\eea
Substituting this yields the simple form
\[
\delta\phi^R_{i,j+1} = \frac{1}{4g} (t_{i,j} - t_{i,j+1})^2.
\]
Since $\delta\phi^A_{i,j+1} = 0$, Peierls’ formula \eqref{eq:pei} gives
\[
  \{ t_{i,j+1}, \, t_{i,j} \}  =  {1\ov 4g} (t_{i,j+1}-t_{i,j})^2 \, .
\]
Since the retarded (advanced) wake lies strictly within the discrete future (past) lightcone, time variables separated spatially on the lattice have zero Poisson brackets.

As a consistency check, consider the flat space limit, where the AdS time variables are parametrized as
\be
   t_{i,j} = \tan  {   t^{(0)}_{i,j} \pm {\pi \ov 2} \ov 2}
\ee
with infinitesimally small $t^{(0)}$. This corresponds to placing the string in a small region of AdS. In this limit, the  equation of motion becomes linear
\[
  t^{(0)}_{i,j+1} =   t^{(0)}_{i-1,j} +   t^{(0)}_{i+1,j} -   t^{(0)}_{i,j-1}
\]
and the Poisson brackets simplify to
\[
\{   t^{(0)}_{i,j+1}, \,   t^{(0)}_{i,j} \}  =  {1\ov g } \, .
\]
Using these results, one can straightforwardly reproduce the Dirac brackets found in \cite{Soederberg}.
We note that in \cite{Soederberg} the dynamical variables are defined along the discrete time direction, which leads to more complicated bracket structures compared to our formulation where the variables lie on two adjacent spatial slices.

\bigskip

\begin{center}
\textbf{Appendix B -- The Casimir}
\end{center}

In global coordinates, the AdS$_2$ metric is
\[
    ds^2 = R^2 (-\cosh^2 \rho \, d\tau^2 + d\rho^2 ) \, ,  
\]
Let us consider a simple folded string with two particles in this target space.
We place the string in the `middle' with endpoints at $\rho = \pm \tilde \rho(\tau)$.
The action can be written as \cite{Ficnar:2013wba}
\begin{align*}
  S = & -{1 \over 4\pi\alpha'} \int  d^2 \sigma \, \sqrt{-h} h^{ab}
    \partial_a X^\mu \partial_b X^\nu G_{\mu\nu}  \qquad\qquad \\
    & \qquad\qquad\qquad\qquad\qquad +
   \int  d\xi \, {1 \over 2\eta} \dot{X}^\mu \dot{X}^\nu G_{\mu\nu} \,,
\end{align*}
where the first integral is the P{olyakov action and the second integral is over the locations of the two particles at the endpoints. Here $X$ is the embedding function, $h$ is the worldsheet metric, $G$ is the target space metric, and $\eta$ is a Lagrange multiplier ensuring that the particles move at the speed of light.

After eliminating $\eta$, the center-of-mass Hamiltonian is the sum of kinetic and potential energies \cite{Callebaut:2015fsa}
\[ 
 H(\tilde p, \tilde \rho) = T+V = |\tilde{p}| \cosh \tilde \rho + {4g} \sinh |\tilde \rho |\, ,
\]
where $\tilde{p}$ is the  momentum conjugate to $\tilde \rho$ and $g = {R^2 \ov 2\pi \alpha'}$. Let us consider a time $\tau_0$ when the two particles reach their turning points. There we have $\tilde{p}(\tau_0)=0$, and the energy equals the potential
\be
  \label{eq:potterm}
 V(\tilde \rho) = {4g} \sinh |\tilde \rho |\, .
\ee
We have $ 2 \tilde \rho = d/R$ where $d$ is the geodesic distance between the two particles at $\rho = \pm \tilde \rho(\tau_0)$.
The geodesic length between turning points can be computed from the $t_{i,j}$ variables. It is simplest to do the calculation in the $\RR^{2,1}$ embedding space of AdS$_2$ as follows.
A pair of retarded and advanced times $(a,b)$ defines a point in $\RR^{2,1}$:
\[
  X(a,b) \equiv \le({1-ab\ov a-b}, \,  {a+b\ov a-b}, \, {1+ab\ov a-b} \ri)   \, ,
\]
which satisfies $X^2 = -1$.
The geodesic distance $d$ between $(t_1, \tilde t_1)$ and $(t_2, \tilde t_2)$ is then given by
\[
  \cosh {d \ov R} = -X(t_1, \tilde t_1) \cdot X(\tilde t_2,  t_2) \, .
\]
This gives $d$ and thus $\tilde \rho$ in terms of the time variables. Hence, \eqref{eq:potterm} yields the cross ratio for the mass-squared
\[  M^2 = V(\tilde \rho)^2 = 16g^2{ (\tilde t_1 - t_2)(\tilde t_2 - t_1) \ov (t_1 - \tilde t_1)(\tilde t_2 - t_2)} \, .
\]
Note that this expression is invariant under the discrete time step \eqref{eq:eom}. For two particles this means that we replace $(t_i, \tilde t_i) \to (\tilde t_i, \tilde{\tilde{t}}_i)$ where
\[
  \tilde{\tilde{t}}_1 = {2 \tilde t_1 \tilde t_2 - t_1( \tilde t_1 +\tilde t_2 ) \ov  \tilde t_1 +\tilde t_2  - 2t_1} \, , \quad
  \tilde{\tilde{t}}_2 = {2 \tilde t_1 \tilde t_2 - t_2( \tilde t_1 +\tilde t_2 ) \ov  \tilde t_1 +\tilde t_2  - 2t_2} \, .
\]

Finally, let us compute the classical solution.
Considering two particles arranged symmetrically the time variables satisfy
\[
  t_1 = -(\tilde t_2)^{-1} \, , \qquad  t_2 = -(\tilde t_1)^{-1} \, .
\]
Taking the Hamiltonian $H= M$ and using the quadratic Poisson brackets \eqref{eq:poissonb2}, Hamilton's equations can be expressed as
\[
  t_1'(t) = {1+ t_1(t)^2 \ov 2} \, , \qquad
  \tilde t_1'(t) = {1+ \tilde t_1(t)^2 \ov 2} \, ,
\]
where $t$ parametrizes the motion in phase space.
The equations are solved by
\[
  t_1(t) = \tan {t-t_0 \ov 2} \, \qquad
  \tilde t_1(t) = \tan {t-\tilde t_0 \ov 2} \, ,
\]
where $t_0$ and $\tilde t_0$ are constants.
If $g$ is imaginary, then $t_1 = (\tilde t_1)^*$. Then, the Cayley transform $z \mapsto {z-i \ov z+i}$ maps the motion into the circular motion in phase space as seen in Figure~\ref{fig:rotat}. Since the Hamiltonian is $SL(2)$ invariant, in a general configuration the two points orbit around their ``center of mass."

\clearpage

\bibliography{paper}

\begin{thebibliography}{34}%
\makeatletter
\providecommand \@ifxundefined [1]{%
 \@ifx{#1\undefined}
}%
\providecommand \@ifnum [1]{%
 \ifnum #1\expandafter \@firstoftwo
 \else \expandafter \@secondoftwo
 \fi
}%
\providecommand \@ifx [1]{%
 \ifx #1\expandafter \@firstoftwo
 \else \expandafter \@secondoftwo
 \fi
}%
\providecommand \natexlab [1]{#1}%
\providecommand \enquote  [1]{``#1''}%
\providecommand \bibnamefont  [1]{#1}%
\providecommand \bibfnamefont [1]{#1}%
\providecommand \citenamefont [1]{#1}%
\providecommand \href@noop [0]{\@secondoftwo}%
\providecommand \href [0]{\begingroup \@sanitize@url \@href}%
\providecommand \@href[1]{\@@startlink{#1}\@@href}%
\providecommand \@@href[1]{\endgroup#1\@@endlink}%
\providecommand \@sanitize@url [0]{\catcode `\\12\catcode `\$12\catcode
  `\&12\catcode `\#12\catcode `\^12\catcode `\_12\catcode `\%12\relax}%
\providecommand \@@startlink[1]{}%
\providecommand \@@endlink[0]{}%
\providecommand \url  [0]{\begingroup\@sanitize@url \@url }%
\providecommand \@url [1]{\endgroup\@href {#1}{\urlprefix }}%
\providecommand \urlprefix  [0]{URL }%
\providecommand \Eprint [0]{\href }%
\providecommand \doibase [0]{https://doi.org/}%
\providecommand \selectlanguage [0]{\@gobble}%
\providecommand \bibinfo  [0]{\@secondoftwo}%
\providecommand \bibfield  [0]{\@secondoftwo}%
\providecommand \translation [1]{[#1]}%
\providecommand \BibitemOpen [0]{}%
\providecommand \bibitemStop [0]{}%
\providecommand \bibitemNoStop [0]{.\EOS\space}%
\providecommand \EOS [0]{\spacefactor3000\relax}%
\providecommand \BibitemShut  [1]{\csname bibitem#1\endcsname}%
\let\auto@bib@innerbib\@empty
\bibitem [{\citenamefont {Sachdev}\ and\ \citenamefont
  {Ye}(1993)}]{1993sachdev}%
  \BibitemOpen
  \bibfield  {author} {\bibinfo {author} {\bibfnamefont {S.}~\bibnamefont
  {Sachdev}}\ and\ \bibinfo {author} {\bibfnamefont {J.}~\bibnamefont {Ye}},\
  }\bibfield  {title} {\bibinfo {title} {Gapless spin-fluid ground state in a
  random quantum heisenberg magnet},\ }\href
  {https://doi.org/10.1103/PhysRevLett.70.3339} {\bibfield  {journal} {\bibinfo
   {journal} {Phys. Rev. Lett.}\ }\textbf {\bibinfo {volume} {70}},\ \bibinfo
  {pages} {3339} (\bibinfo {year} {1993})},\ \Eprint
  {https://arxiv.org/abs/cond-mat/9212030} {arXiv:cond-mat/9212030 [cond-mat]}
  \BibitemShut {NoStop}%
\bibitem [{\citenamefont {Kitaev}()}]{kitaev}%
  \BibitemOpen
  \bibfield  {author} {\bibinfo {author} {\bibfnamefont {A.}~\bibnamefont
  {Kitaev}},\ }\bibfield  {title} {\bibinfo {title} {{A Simple Model of Quantum
  Holography, Talks at KITP, April 7, 2015 and May 27, 2015,
  http://online.kitp.ucsb.edu/online/entangled15/kitaev}},\ }\href@noop {} {\
  }\BibitemShut {NoStop}%
\bibitem [{\citenamefont {Polchinski}\ and\ \citenamefont
  {Rosenhaus}(2016)}]{Polchinski:2016xgd}%
  \BibitemOpen
  \bibfield  {author} {\bibinfo {author} {\bibfnamefont {J.}~\bibnamefont
  {Polchinski}}\ and\ \bibinfo {author} {\bibfnamefont {V.}~\bibnamefont
  {Rosenhaus}},\ }\bibfield  {title} {\bibinfo {title} {{The Spectrum in the
  Sachdev-Ye-Kitaev Model}},\ }\href {https://doi.org/10.1007/JHEP04(2016)001}
  {\bibfield  {journal} {\bibinfo  {journal} {JHEP}\ }\textbf {\bibinfo
  {volume} {04}},\ \bibinfo {pages} {001}},\ \Eprint
  {https://arxiv.org/abs/1601.06768} {arXiv:1601.06768 [hep-th]} \BibitemShut
  {NoStop}%
\bibitem [{\citenamefont {Maldacena}\ and\ \citenamefont
  {Stanford}(2016)}]{Maldacena:2016hyu}%
  \BibitemOpen
  \bibfield  {author} {\bibinfo {author} {\bibfnamefont {J.}~\bibnamefont
  {Maldacena}}\ and\ \bibinfo {author} {\bibfnamefont {D.}~\bibnamefont
  {Stanford}},\ }\bibfield  {title} {\bibinfo {title} {{Remarks on the
  Sachdev-Ye-Kitaev model}},\ }\href
  {https://doi.org/10.1103/PhysRevD.94.106002} {\bibfield  {journal} {\bibinfo
  {journal} {Phys. Rev. D}\ }\textbf {\bibinfo {volume} {94}},\ \bibinfo
  {pages} {106002} (\bibinfo {year} {2016})},\ \Eprint
  {https://arxiv.org/abs/1604.07818} {arXiv:1604.07818 [hep-th]} \BibitemShut
  {NoStop}%
\bibitem [{\citenamefont {Jensen}(2016)}]{Jensen:2016pah}%
  \BibitemOpen
  \bibfield  {author} {\bibinfo {author} {\bibfnamefont {K.}~\bibnamefont
  {Jensen}},\ }\bibfield  {title} {\bibinfo {title} {{Chaos in AdS$_2$
  Holography}},\ }\href {https://doi.org/10.1103/PhysRevLett.117.111601}
  {\bibfield  {journal} {\bibinfo  {journal} {Phys. Rev. Lett.}\ }\textbf
  {\bibinfo {volume} {117}},\ \bibinfo {pages} {111601} (\bibinfo {year}
  {2016})},\ \Eprint {https://arxiv.org/abs/1605.06098} {arXiv:1605.06098
  [hep-th]} \BibitemShut {NoStop}%
\bibitem [{\citenamefont {Maldacena}\ \emph {et~al.}(2016)\citenamefont
  {Maldacena}, \citenamefont {Stanford},\ and\ \citenamefont
  {Yang}}]{Maldacena:2016upp}%
  \BibitemOpen
  \bibfield  {author} {\bibinfo {author} {\bibfnamefont {J.}~\bibnamefont
  {Maldacena}}, \bibinfo {author} {\bibfnamefont {D.}~\bibnamefont
  {Stanford}},\ and\ \bibinfo {author} {\bibfnamefont {Z.}~\bibnamefont
  {Yang}},\ }\bibfield  {title} {\bibinfo {title} {{Conformal symmetry and its
  breaking in two dimensional Nearly Anti-de-Sitter space}},\ }\href
  {https://doi.org/10.1093/ptep/ptw124} {\bibfield  {journal} {\bibinfo
  {journal} {PTEP}\ }\textbf {\bibinfo {volume} {2016}},\ \bibinfo {pages}
  {12C104} (\bibinfo {year} {2016})},\ \Eprint
  {https://arxiv.org/abs/1606.01857} {arXiv:1606.01857 [hep-th]} \BibitemShut
  {NoStop}%
\bibitem [{\citenamefont {Soederberg}\ \emph {et~al.}(1985)\citenamefont
  {Soederberg}, \citenamefont {Andersson},\ and\ \citenamefont
  {Gustafson}}]{Soederberg}%
  \BibitemOpen
  \bibfield  {author} {\bibinfo {author} {\bibfnamefont {B.}~\bibnamefont
  {Soederberg}}, \bibinfo {author} {\bibfnamefont {B.}~\bibnamefont
  {Andersson}},\ and\ \bibinfo {author} {\bibfnamefont {G.}~\bibnamefont
  {Gustafson}},\ }\bibfield  {title} {\bibinfo {title} {{Action-angle variables
  for the massless relativistic string in 1+1 dimensions}},\ }\href@noop {}
  {\bibfield  {journal} {\bibinfo  {journal} {J Math Phys (NY)}\ }\textbf
  {\bibinfo {volume} {26(1)}},\ \bibinfo {pages} {112} (\bibinfo {year}
  {1985})}\BibitemShut {NoStop}%
\bibitem [{\citenamefont {Bars}\ and\ \citenamefont
  {Schulze}(1995)}]{Bars:1994sv}%
  \BibitemOpen
  \bibfield  {author} {\bibinfo {author} {\bibfnamefont {I.}~\bibnamefont
  {Bars}}\ and\ \bibinfo {author} {\bibfnamefont {J.}~\bibnamefont {Schulze}},\
  }\bibfield  {title} {\bibinfo {title} {{Folded strings falling into a black
  hole}},\ }\href {https://doi.org/10.1103/PhysRevD.51.1854} {\bibfield
  {journal} {\bibinfo  {journal} {Phys. Rev. D}\ }\textbf {\bibinfo {volume}
  {51}},\ \bibinfo {pages} {1854} (\bibinfo {year} {1995})},\ \Eprint
  {https://arxiv.org/abs/hep-th/9405156} {arXiv:hep-th/9405156} \BibitemShut
  {NoStop}%
\bibitem [{\citenamefont {Bars}(1994)}]{Bars:1994xi}%
  \BibitemOpen
  \bibfield  {author} {\bibinfo {author} {\bibfnamefont {I.}~\bibnamefont
  {Bars}},\ }\bibfield  {title} {\bibinfo {title} {{Folded strings in curved
  space-time}},\ }\href@noop {} {\  (\bibinfo {year} {1994})},\ \Eprint
  {https://arxiv.org/abs/hep-th/9411078} {arXiv:hep-th/9411078} \BibitemShut
  {NoStop}%
\bibitem [{\citenamefont {Vegh}(2016)}]{Vegh:2016hwq}%
  \BibitemOpen
  \bibfield  {author} {\bibinfo {author} {\bibfnamefont {D.}~\bibnamefont
  {Vegh}},\ }\bibfield  {title} {\bibinfo {title} {{Segmented strings from a
  different angle}},\ }\href@noop {} {\  (\bibinfo {year} {2016})},\ \Eprint
  {https://arxiv.org/abs/1601.07571} {arXiv:1601.07571 [hep-th]} \BibitemShut
  {NoStop}%
\bibitem [{\citenamefont {Vegh}(2018)}]{Vegh:2016fcm}%
  \BibitemOpen
  \bibfield  {author} {\bibinfo {author} {\bibfnamefont {D.}~\bibnamefont
  {Vegh}},\ }\bibfield  {title} {\bibinfo {title} {{Segmented strings coupled
  to a B-field}},\ }\href {https://doi.org/10.1007/JHEP04(2018)088} {\bibfield
  {journal} {\bibinfo  {journal} {JHEP}\ }\textbf {\bibinfo {volume} {04}},\
  \bibinfo {pages} {088}},\ \Eprint {https://arxiv.org/abs/1603.04504}
  {arXiv:1603.04504 [hep-th]} \BibitemShut {NoStop}%
\bibitem [{\citenamefont {Peierls}(1952)}]{Peierls:1952cb}%
  \BibitemOpen
  \bibfield  {author} {\bibinfo {author} {\bibfnamefont {R.~E.}\ \bibnamefont
  {Peierls}},\ }\bibfield  {title} {\bibinfo {title} {{The Commutation laws of
  relativistic field theory}},\ }\href {https://doi.org/10.1098/rspa.1952.0158}
  {\bibfield  {journal} {\bibinfo  {journal} {Proc. Roy. Soc. Lond. A}\
  }\textbf {\bibinfo {volume} {214}},\ \bibinfo {pages} {143} (\bibinfo {year}
  {1952})}\BibitemShut {NoStop}%
\bibitem [{\citenamefont {Gekhtman}\ \emph {et~al.}(2016)\citenamefont
  {Gekhtman}, \citenamefont {Shapiro}, \citenamefont {Tabachnikov},\ and\
  \citenamefont {Vainshtein}}]{GEKHTMAN2016390}%
  \BibitemOpen
  \bibfield  {author} {\bibinfo {author} {\bibfnamefont {M.}~\bibnamefont
  {Gekhtman}}, \bibinfo {author} {\bibfnamefont {M.}~\bibnamefont {Shapiro}},
  \bibinfo {author} {\bibfnamefont {S.}~\bibnamefont {Tabachnikov}},\ and\
  \bibinfo {author} {\bibfnamefont {A.}~\bibnamefont {Vainshtein}},\ }\bibfield
   {title} {\bibinfo {title} {Integrable cluster dynamics of directed networks
  and pentagram maps},\ }\href
  {https://doi.org/https://doi.org/10.1016/j.aim.2016.03.023} {\bibfield
  {journal} {\bibinfo  {journal} {Advances in Mathematics}\ }\textbf {\bibinfo
  {volume} {300}},\ \bibinfo {pages} {390} (\bibinfo {year} {2016})},\ \bibinfo
  {note} {special volume honoring Andrei Zelevinsky}\BibitemShut {NoStop}%
\bibitem [{\citenamefont {Lenz}\ and\ \citenamefont
  {Schreiber}(1996)}]{Lenz:1995tj}%
  \BibitemOpen
  \bibfield  {author} {\bibinfo {author} {\bibfnamefont {S.}~\bibnamefont
  {Lenz}}\ and\ \bibinfo {author} {\bibfnamefont {B.}~\bibnamefont
  {Schreiber}},\ }\bibfield  {title} {\bibinfo {title} {{Example of a Poincare
  anomaly in relativistic quantum mechanics}},\ }\href
  {https://doi.org/10.1103/PhysRevD.53.960} {\bibfield  {journal} {\bibinfo
  {journal} {Phys. Rev. D}\ }\textbf {\bibinfo {volume} {53}},\ \bibinfo
  {pages} {960} (\bibinfo {year} {1996})},\ \Eprint
  {https://arxiv.org/abs/hep-th/9503219} {arXiv:hep-th/9503219} \BibitemShut
  {NoStop}%
\bibitem [{\citenamefont {'t~Hooft}(1974)}]{tHooft:1974pnl}%
  \BibitemOpen
  \bibfield  {author} {\bibinfo {author} {\bibfnamefont {G.}~\bibnamefont
  {'t~Hooft}},\ }\bibfield  {title} {\bibinfo {title} {{A Two-Dimensional Model
  for Mesons}},\ }\href {https://doi.org/10.1016/0550-3213(74)90088-1}
  {\bibfield  {journal} {\bibinfo  {journal} {Nucl. Phys. B}\ }\textbf
  {\bibinfo {volume} {75}},\ \bibinfo {pages} {461} (\bibinfo {year}
  {1974})}\BibitemShut {NoStop}%
\bibitem [{\citenamefont {Suris}(2003)}]{suris}%
  \BibitemOpen
  \bibfield  {author} {\bibinfo {author} {\bibfnamefont {Y.~B.}\ \bibnamefont
  {Suris}},\ }\href@noop {} {\emph {\bibinfo {title} {The Problem of Integrable
  Discretization: Hamiltonian Approach}}}\ (\bibinfo  {publisher} {Birkh\"
  auser Verlag},\ \bibinfo {address} {Basel},\ \bibinfo {year}
  {2003})\BibitemShut {NoStop}%
\bibitem [{\citenamefont {Vegh}(2023)}]{Vegh:2023snc}%
  \BibitemOpen
  \bibfield  {author} {\bibinfo {author} {\bibfnamefont {D.}~\bibnamefont
  {Vegh}},\ }\bibfield  {title} {\bibinfo {title} {{The 't Hooft equation as a
  quantum spectral curve}},\ }\href@noop {} {\  (\bibinfo {year} {2023})},\
  \Eprint {https://arxiv.org/abs/2301.07154} {arXiv:2301.07154 [hep-th]}
  \BibitemShut {NoStop}%
\bibitem [{\citenamefont {Stanford}\ and\ \citenamefont
  {Witten}(2017)}]{Stanford:2017thb}%
  \BibitemOpen
  \bibfield  {author} {\bibinfo {author} {\bibfnamefont {D.}~\bibnamefont
  {Stanford}}\ and\ \bibinfo {author} {\bibfnamefont {E.}~\bibnamefont
  {Witten}},\ }\bibfield  {title} {\bibinfo {title} {{Fermionic Localization of
  the Schwarzian Theory}},\ }\href {https://doi.org/10.1007/JHEP10(2017)008}
  {\bibfield  {journal} {\bibinfo  {journal} {JHEP}\ }\textbf {\bibinfo
  {volume} {10}},\ \bibinfo {pages} {008}},\ \Eprint
  {https://arxiv.org/abs/1703.04612} {arXiv:1703.04612 [hep-th]} \BibitemShut
  {NoStop}%
\bibitem [{\citenamefont {Vegh}(2024)}]{Vegh:2024uie}%
  \BibitemOpen
  \bibfield  {author} {\bibinfo {author} {\bibfnamefont {D.}~\bibnamefont
  {Vegh}},\ }\bibfield  {title} {\bibinfo {title} {{Quantizing the folded
  string in AdS$_2$}},\ }\href@noop {} {\  (\bibinfo {year} {2024})},\ \Eprint
  {https://arxiv.org/abs/2409.06663} {arXiv:2409.06663 [hep-th]} \BibitemShut
  {NoStop}%
\bibitem [{\citenamefont {Poschl}\ and\ \citenamefont
  {Teller}(1933)}]{Poschl:1933zz}%
  \BibitemOpen
  \bibfield  {author} {\bibinfo {author} {\bibfnamefont {G.}~\bibnamefont
  {Poschl}}\ and\ \bibinfo {author} {\bibfnamefont {E.}~\bibnamefont
  {Teller}},\ }\bibfield  {title} {\bibinfo {title} {{Bemerkungen zur
  Quantenmechanik des anharmonischen Oszillators}},\ }\href
  {https://doi.org/10.1007/BF01331132} {\bibfield  {journal} {\bibinfo
  {journal} {Z. Phys.}\ }\textbf {\bibinfo {volume} {83}},\ \bibinfo {pages}
  {143} (\bibinfo {year} {1933})}\BibitemShut {NoStop}%
\bibitem [{\citenamefont {Gross}\ and\ \citenamefont
  {Rosenhaus}(2017)}]{Gross:2017vhb}%
  \BibitemOpen
  \bibfield  {author} {\bibinfo {author} {\bibfnamefont {D.~J.}\ \bibnamefont
  {Gross}}\ and\ \bibinfo {author} {\bibfnamefont {V.}~\bibnamefont
  {Rosenhaus}},\ }\bibfield  {title} {\bibinfo {title} {{A line of CFTs: from
  generalized free fields to SYK}},\ }\href
  {https://doi.org/10.1007/JHEP07(2017)086} {\bibfield  {journal} {\bibinfo
  {journal} {JHEP}\ }\textbf {\bibinfo {volume} {07}},\ \bibinfo {pages}
  {086}},\ \Eprint {https://arxiv.org/abs/1706.07015} {arXiv:1706.07015
  [hep-th]} \BibitemShut {NoStop}%
\bibitem [{\citenamefont {Anninos}\ \emph {et~al.}(2024)\citenamefont
  {Anninos}, \citenamefont {Anous}, \citenamefont {Pethybridge},\ and\
  \citenamefont {{\c{S}}eng{\"o}r}}]{Anninos:2023lin}%
  \BibitemOpen
  \bibfield  {author} {\bibinfo {author} {\bibfnamefont {D.}~\bibnamefont
  {Anninos}}, \bibinfo {author} {\bibfnamefont {T.}~\bibnamefont {Anous}},
  \bibinfo {author} {\bibfnamefont {B.}~\bibnamefont {Pethybridge}},\ and\
  \bibinfo {author} {\bibfnamefont {G.}~\bibnamefont {{\c{S}}eng{\"o}r}},\
  }\bibfield  {title} {\bibinfo {title} {{The discreet charm of the discrete
  series in dS$_{2}$}},\ }\href {https://doi.org/10.1088/1751-8121/ad14ad}
  {\bibfield  {journal} {\bibinfo  {journal} {J. Phys. A}\ }\textbf {\bibinfo
  {volume} {57}},\ \bibinfo {pages} {025401} (\bibinfo {year} {2024})},\
  \Eprint {https://arxiv.org/abs/2307.15832} {arXiv:2307.15832 [hep-th]}
  \BibitemShut {NoStop}%
\bibitem [{\citenamefont {Pethybridge}(2024)}]{Pethybridge:2024qci}%
  \BibitemOpen
  \bibfield  {author} {\bibinfo {author} {\bibfnamefont {B.~J.}\ \bibnamefont
  {Pethybridge}},\ }\bibfield  {title} {\bibinfo {title} {{Notes on complex
  $q=2$ SYK}},\ }\href@noop {} {\  (\bibinfo {year} {2024})},\ \Eprint
  {https://arxiv.org/abs/2403.04673} {arXiv:2403.04673 [hep-th]} \BibitemShut
  {NoStop}%
\bibitem [{\citenamefont {Zamolodchikov}(1980)}]{Zamolodchikov:1980mb}%
  \BibitemOpen
  \bibfield  {author} {\bibinfo {author} {\bibfnamefont {A.~B.}\ \bibnamefont
  {Zamolodchikov}},\ }\bibfield  {title} {\bibinfo {title} {{'Fishnet' diagrams
  as a completely integrable system}},\ }\href
  {https://doi.org/10.1016/0370-2693(80)90547-X} {\bibfield  {journal}
  {\bibinfo  {journal} {Phys. Lett. B}\ }\textbf {\bibinfo {volume} {97}},\
  \bibinfo {pages} {63} (\bibinfo {year} {1980})}\BibitemShut {NoStop}%
\bibitem [{\citenamefont {Kazakov}\ and\ \citenamefont
  {Olivucci}(2018)}]{Kazakov:2018qbr}%
  \BibitemOpen
  \bibfield  {author} {\bibinfo {author} {\bibfnamefont {V.}~\bibnamefont
  {Kazakov}}\ and\ \bibinfo {author} {\bibfnamefont {E.}~\bibnamefont
  {Olivucci}},\ }\bibfield  {title} {\bibinfo {title} {{Biscalar Integrable
  Conformal Field Theories in Any Dimension}},\ }\href
  {https://doi.org/10.1103/PhysRevLett.121.131601} {\bibfield  {journal}
  {\bibinfo  {journal} {Phys. Rev. Lett.}\ }\textbf {\bibinfo {volume} {121}},\
  \bibinfo {pages} {131601} (\bibinfo {year} {2018})},\ \Eprint
  {https://arxiv.org/abs/1801.09844} {arXiv:1801.09844 [hep-th]} \BibitemShut
  {NoStop}%
\bibitem [{\citenamefont {Kazakov}\ and\ \citenamefont
  {Olivucci}(2023)}]{Kazakov:2022dbd}%
  \BibitemOpen
  \bibfield  {author} {\bibinfo {author} {\bibfnamefont {V.}~\bibnamefont
  {Kazakov}}\ and\ \bibinfo {author} {\bibfnamefont {E.}~\bibnamefont
  {Olivucci}},\ }\bibfield  {title} {\bibinfo {title} {{The loom for general
  fishnet CFTs}},\ }\href {https://doi.org/10.1007/JHEP06(2023)041} {\bibfield
  {journal} {\bibinfo  {journal} {JHEP}\ }\textbf {\bibinfo {volume} {06}},\
  \bibinfo {pages} {041}},\ \Eprint {https://arxiv.org/abs/2212.09732}
  {arXiv:2212.09732 [hep-th]} \BibitemShut {NoStop}%
\bibitem [{\citenamefont {Kazakov}\ \emph {et~al.}(2025)\citenamefont
  {Kazakov}, \citenamefont {Levkovich-Maslyuk},\ and\ \citenamefont
  {Mishnyakov}}]{Kazakov:2023nyu}%
  \BibitemOpen
  \bibfield  {author} {\bibinfo {author} {\bibfnamefont {V.}~\bibnamefont
  {Kazakov}}, \bibinfo {author} {\bibfnamefont {F.}~\bibnamefont
  {Levkovich-Maslyuk}},\ and\ \bibinfo {author} {\bibfnamefont
  {V.}~\bibnamefont {Mishnyakov}},\ }\bibfield  {title} {\bibinfo {title}
  {{Integrable Feynman graphs and Yangian symmetry on the loom}},\ }\href
  {https://doi.org/10.1007/JHEP06(2025)104} {\bibfield  {journal} {\bibinfo
  {journal} {JHEP}\ }\textbf {\bibinfo {volume} {06}},\ \bibinfo {pages}
  {104}},\ \Eprint {https://arxiv.org/abs/2304.04654} {arXiv:2304.04654
  [hep-th]} \BibitemShut {NoStop}%
\bibitem [{\citenamefont {Gromov}\ and\ \citenamefont
  {Sever}(2019{\natexlab{a}})}]{Gromov:2019aku}%
  \BibitemOpen
  \bibfield  {author} {\bibinfo {author} {\bibfnamefont {N.}~\bibnamefont
  {Gromov}}\ and\ \bibinfo {author} {\bibfnamefont {A.}~\bibnamefont {Sever}},\
  }\bibfield  {title} {\bibinfo {title} {{Derivation of the Holographic Dual of
  a Planar Conformal Field Theory in 4D}},\ }\href
  {https://doi.org/10.1103/PhysRevLett.123.081602} {\bibfield  {journal}
  {\bibinfo  {journal} {Phys. Rev. Lett.}\ }\textbf {\bibinfo {volume} {123}},\
  \bibinfo {pages} {081602} (\bibinfo {year} {2019}{\natexlab{a}})},\ \Eprint
  {https://arxiv.org/abs/1903.10508} {arXiv:1903.10508 [hep-th]} \BibitemShut
  {NoStop}%
\bibitem [{\citenamefont {Gromov}\ and\ \citenamefont
  {Sever}(2019{\natexlab{b}})}]{Gromov:2019bsj}%
  \BibitemOpen
  \bibfield  {author} {\bibinfo {author} {\bibfnamefont {N.}~\bibnamefont
  {Gromov}}\ and\ \bibinfo {author} {\bibfnamefont {A.}~\bibnamefont {Sever}},\
  }\bibfield  {title} {\bibinfo {title} {{Quantum fishchain in AdS$_{5}$}},\
  }\href {https://doi.org/10.1007/JHEP10(2019)085} {\bibfield  {journal}
  {\bibinfo  {journal} {JHEP}\ }\textbf {\bibinfo {volume} {10}},\ \bibinfo
  {pages} {085}},\ \Eprint {https://arxiv.org/abs/1907.01001} {arXiv:1907.01001
  [hep-th]} \BibitemShut {NoStop}%
\bibitem [{\citenamefont {Gromov}\ and\ \citenamefont
  {Sever}(2020)}]{Gromov:2019jfh}%
  \BibitemOpen
  \bibfield  {author} {\bibinfo {author} {\bibfnamefont {N.}~\bibnamefont
  {Gromov}}\ and\ \bibinfo {author} {\bibfnamefont {A.}~\bibnamefont {Sever}},\
  }\bibfield  {title} {\bibinfo {title} {{The holographic dual of strongly
  $\gamma$-deformed $ \mathcal{N} $ = 4 SYM theory: derivation, generalization,
  integrability and discrete reparametrization symmetry}},\ }\href
  {https://doi.org/10.1007/JHEP02(2020)035} {\bibfield  {journal} {\bibinfo
  {journal} {JHEP}\ }\textbf {\bibinfo {volume} {02}},\ \bibinfo {pages}
  {035}},\ \Eprint {https://arxiv.org/abs/1908.10379} {arXiv:1908.10379
  [hep-th]} \BibitemShut {NoStop}%
\bibitem [{\citenamefont {Gu}\ \emph {et~al.}(2020)\citenamefont {Gu},
  \citenamefont {Kitaev}, \citenamefont {Sachdev},\ and\ \citenamefont
  {Tarnopolsky}}]{Gu:2019jub}%
  \BibitemOpen
  \bibfield  {author} {\bibinfo {author} {\bibfnamefont {Y.}~\bibnamefont
  {Gu}}, \bibinfo {author} {\bibfnamefont {A.}~\bibnamefont {Kitaev}}, \bibinfo
  {author} {\bibfnamefont {S.}~\bibnamefont {Sachdev}},\ and\ \bibinfo {author}
  {\bibfnamefont {G.}~\bibnamefont {Tarnopolsky}},\ }\bibfield  {title}
  {\bibinfo {title} {{Notes on the complex Sachdev-Ye-Kitaev model}},\ }\href
  {https://doi.org/10.1007/JHEP02(2020)157} {\bibfield  {journal} {\bibinfo
  {journal} {JHEP}\ }\textbf {\bibinfo {volume} {02}},\ \bibinfo {pages}
  {157}},\ \Eprint {https://arxiv.org/abs/1910.14099} {arXiv:1910.14099
  [hep-th]} \BibitemShut {NoStop}%
\bibitem [{\citenamefont {Fu}\ \emph {et~al.}(2017)\citenamefont {Fu},
  \citenamefont {Gaiotto}, \citenamefont {Maldacena},\ and\ \citenamefont
  {Sachdev}}]{Fu:2016vas}%
  \BibitemOpen
  \bibfield  {author} {\bibinfo {author} {\bibfnamefont {W.}~\bibnamefont
  {Fu}}, \bibinfo {author} {\bibfnamefont {D.}~\bibnamefont {Gaiotto}},
  \bibinfo {author} {\bibfnamefont {J.}~\bibnamefont {Maldacena}},\ and\
  \bibinfo {author} {\bibfnamefont {S.}~\bibnamefont {Sachdev}},\ }\bibfield
  {title} {\bibinfo {title} {{Supersymmetric Sachdev-Ye-Kitaev models}},\
  }\href {https://doi.org/10.1103/PhysRevD.95.026009} {\bibfield  {journal}
  {\bibinfo  {journal} {Phys. Rev. D}\ }\textbf {\bibinfo {volume} {95}},\
  \bibinfo {pages} {026009} (\bibinfo {year} {2017})},\ \bibinfo {note}
  {[Addendum: Phys.Rev.D 95, 069904 (2017)]},\ \Eprint
  {https://arxiv.org/abs/1610.08917} {arXiv:1610.08917 [hep-th]} \BibitemShut
  {NoStop}%
\bibitem [{\citenamefont {Ficnar}\ and\ \citenamefont
  {Gubser}(2014)}]{Ficnar:2013wba}%
  \BibitemOpen
  \bibfield  {author} {\bibinfo {author} {\bibfnamefont {A.}~\bibnamefont
  {Ficnar}}\ and\ \bibinfo {author} {\bibfnamefont {S.~S.}\ \bibnamefont
  {Gubser}},\ }\bibfield  {title} {\bibinfo {title} {{Finite momentum at string
  endpoints}},\ }\href {https://doi.org/10.1103/PhysRevD.89.026002} {\bibfield
  {journal} {\bibinfo  {journal} {Phys. Rev.}\ }\textbf {\bibinfo {volume}
  {D89}},\ \bibinfo {pages} {026002} (\bibinfo {year} {2014})},\ \Eprint
  {https://arxiv.org/abs/1306.6648} {arXiv:1306.6648 [hep-th]} \BibitemShut
  {NoStop}%
\bibitem [{\citenamefont {Callebaut}\ \emph {et~al.}(2015)\citenamefont
  {Callebaut}, \citenamefont {Gubser}, \citenamefont {Samberg},\ and\
  \citenamefont {Toldo}}]{Callebaut:2015fsa}%
  \BibitemOpen
  \bibfield  {author} {\bibinfo {author} {\bibfnamefont {N.}~\bibnamefont
  {Callebaut}}, \bibinfo {author} {\bibfnamefont {S.~S.}\ \bibnamefont
  {Gubser}}, \bibinfo {author} {\bibfnamefont {A.}~\bibnamefont {Samberg}},\
  and\ \bibinfo {author} {\bibfnamefont {C.}~\bibnamefont {Toldo}},\ }\bibfield
   {title} {\bibinfo {title} {{Segmented strings in AdS$_{3}$}},\ }\href
  {https://doi.org/10.1007/JHEP11(2015)110} {\bibfield  {journal} {\bibinfo
  {journal} {JHEP}\ }\textbf {\bibinfo {volume} {11}},\ \bibinfo {pages}
  {110}},\ \Eprint {https://arxiv.org/abs/1508.07311} {arXiv:1508.07311
  [hep-th]} \BibitemShut {NoStop}%
\end{thebibliography}%

\end{document}